\documentclass[fleqn,usenatbib]{mnras}

\usepackage{newtxtext,newtxmath}

\usepackage[T1]{fontenc}
\usepackage{ae,aecompl}

\usepackage{graphicx}	\usepackage{amsmath}	\usepackage{bm}
\usepackage{booktabs}
\usepackage{natbib}
\defcitealias{Peng:2015}{P15}
\defcitealias{Trussler:2020}{T20}

\newcommand{\ppxf}{\texttt{pPXF}}

\newcommand{\Msun}{M$_{\odot}$}
\newcommand{\lmstar}{$\log_{10}(\mathrm{M}_{*} / \mathrm{M}_{\odot})$}

\DeclareRobustCommand{\ion}[2]{\relax\ifmmode
\ifx\testbx\f@series
{\mathbf{#1\,\mathsc{#2}}}\else
{\mathrm{#1\,\mathsc{#2}}}\fi
\else\textup{#1\,{\mdseries\textsc{#2}}}\fi}

\title[Mass, Size and Stellar Metallicity]{The SAMI galaxy survey: galaxy size can explain the offset between star-forming and passive galaxies in the mass-metallicity relationship}

\author[S. P. Vaughan et al.]{Sam P. Vaughan$^{1,2}$\thanks{Contact e-mail: \href{mailto:sam.vaughan@sydney.edu.au}{sam.vaughan@sydney.edu.au} (SPV)}, Tania M. Barone$^{1,2,3,4}$, Scott M. Croom$^{1,2}$, Luca Cortese$^{2,5}$, Francesco D'Eugenio$^{6,7}$, 
\newauthor Sarah Brough$^{2, 8}$, Matthew Colless$^{2,4}$, Richard M. McDermid$^{2,9}$,  Jesse van de Sande$^{1,2}$, 
\newauthor Nicholas Scott$^{1,2}$, Joss Bland-Hawthorn$^{1,2}$, Julia J. Bryant$^{1,2}$, J.S. Lawrence$^{12}$, \'Angel R. L\'opez-S\'anchez$^{2,9,11}$,  
\newauthor Nuria P. F. Lorente$^{13}$, Matt S. Owers$^{2,9,10}$ and Samuel N. Richards$^{1}$
\\
$^{1}$Sydney Institute for Astronomy, School of Physics, Building A28, The University of Sydney, NSW 2006, Australia\\
$^{2}$ARC Centre of Excellence for All Sky Astrophysics in 3 Dimensions (ASTRO3D), Australia\\
$^{3}$Centre for Astrophysics and Supercomputing, School of Science, Swinburne University of Technology,
Hawthorn, VIC 3122, Australia.\\
$^{4}$Research School of Astronomy and Astrophysics, Australian National University, Canberra, ACT 2611, Australia\\
$^{5}$ International Centre for Radio Astronomy Research, The University of Western Australia, 35 Stirling Hw, 6009 Crawley, Australia\\
$^{6}$Cavendish Laboratory and Kavli Institute for Cosmology, University of Cambridge, Madingley Rise, Cambridge, CB3 0HA, United Kingdom\\
$^{7}$Sterrenkundig Observatorium, Universiteit Gent, Krijgslaan 281 S9, B-9000 Gent, Belgium\\
$^{8}$School of Physics, University of New South Wales, NSW 2052, Australia\\
$^{9}$Department of Physics and Astronomy, Macquarie University, NSW 2109, Australia\\
$^{10}$Astronomy, Astrophysics and Astrophotonics Research Centre, Macquarie University, Sydney, NSW 2109, Australia\\
$^{11}$Australian Astronomical Optics, Macquarie University, 105 Delhi Rd, North Ryde, NSW 2113, Australia\\
$^{12}$Australian Astronomical Optics - Macquarie, Macquarie University, NSW 2109, Australia\\
$^{13}$AAO-MQ, Faculty of Science \& Engineering, Macquarie University, 105 Delhi Rd, North Ryde, NSW 2113, Australia\\
}
\date{Accepted 2022 August 12. Received 2022 August 12; in original form 2021 November 14}

\pubyear{2019}

\begin{document}
\label{firstpage}
\pagerange{\pageref{firstpage}--\pageref{lastpage}}
\maketitle

\begin{abstract}
In this work, we investigate how the central stellar metallicity ([Z/H]) of 1363 galaxies from the SAMI galaxy survey is related to their stellar mass and a proxy for the gravitational potential, $\Phi = \log_{10}\left(\frac{M_*}{M_{\odot}} \right) - \log_{10}\left(\frac{r_e}{\mathrm{kpc}} \right)$. In agreement with previous studies, we find that passive and star-forming galaxies occupy different areas of the [Z/H]-$\mathrm{M}_{*}$ plane, with passive galaxies having higher [Z/H] than star-forming galaxies at fixed mass (a difference of 0.23 dex at \lmstar=10.3). We show for the first time that all galaxies lie on the same relation between [Z/H] and $\Phi$, and show that the offset in [Z/H] between passive and star-forming galaxies at fixed $\Phi$ is smaller than or equal to the offset in [Z/H] at fixed mass (an average $\Delta$[Z/H] of 0.11 dex at fixed $\Phi$ compared to 0.21 dex at fixed mass). We then build a simple model of galaxy evolution to explain and understand our results. By assuming that [Z/H] traces $\Phi$ over cosmic time and that the probability that a galaxy quenches depends on both its mass and size, we are able to reproduce these offsets in stellar metallicity with a model containing instantaneous quenching. We therefore conclude that an offset in metallicity at fixed mass cannot by itself be used as evidence of slow quenching processes, in contrast to previous studies. Instead, our model implies that metal-rich galaxies have always been the smallest objects for their mass in a
population. Our findings reiterate the need to consider galaxy size when studying stellar populations.
\end{abstract}

\begin{keywords}
galaxies: abundances -- galaxies: evolution --
galaxies: formation 
\end{keywords}

\section{Introduction}

Galaxies in the Universe can broadly be classified into two categories: those which are currently forming stars and those which are not. Understanding and quantifying exactly which physical processes play a role in shaping this dichotomy is a fundamental goal of galaxy evolution, but despite decades of work on the topic \citep[e.g.][among many others]{Tinsley:1968,Larson:1980, Dekel:2006, Peng:2010, Schawinski:2014} a complete picture remains out of reach. 

In the most popular framework for discussing the life cycles of galaxies, the idea of star formation "quenching" plays a central role. This model envisions that the natural state of a galaxy is to be forming stars at a rate which places it on the "main-sequence" of star-formation \citep{Noeske:2007, Speagle:2014,Renzini:2015}. Internal or external processes then act to stop the galaxy's star formation, causing it to transition off the main sequence and join the passive population. The question of "what leads to the division in galaxy properties?" can then be reframed as "what causes quenching?". A wide range of physical drivers of quenching have been proposed and investigated, including feedback from a galaxy's central black hole \citep{Silk:1998, Magorrian:1998, DiMatteo:2005}, the effects of falling into a massive galaxy cluster \citep{Gunn:1972, Abadi:1999}, the secular evolution of stellar orbits over cosmic time \citep{Kormendy:2004, Sellwood:2014} and the occurrence of "compaction" events which increase a galaxy's central surface density \citep[e.g.][]{Woo:2015, Woo:2019}. 

\begin{figure*}
    \centering
    \includegraphics[width=\textwidth]{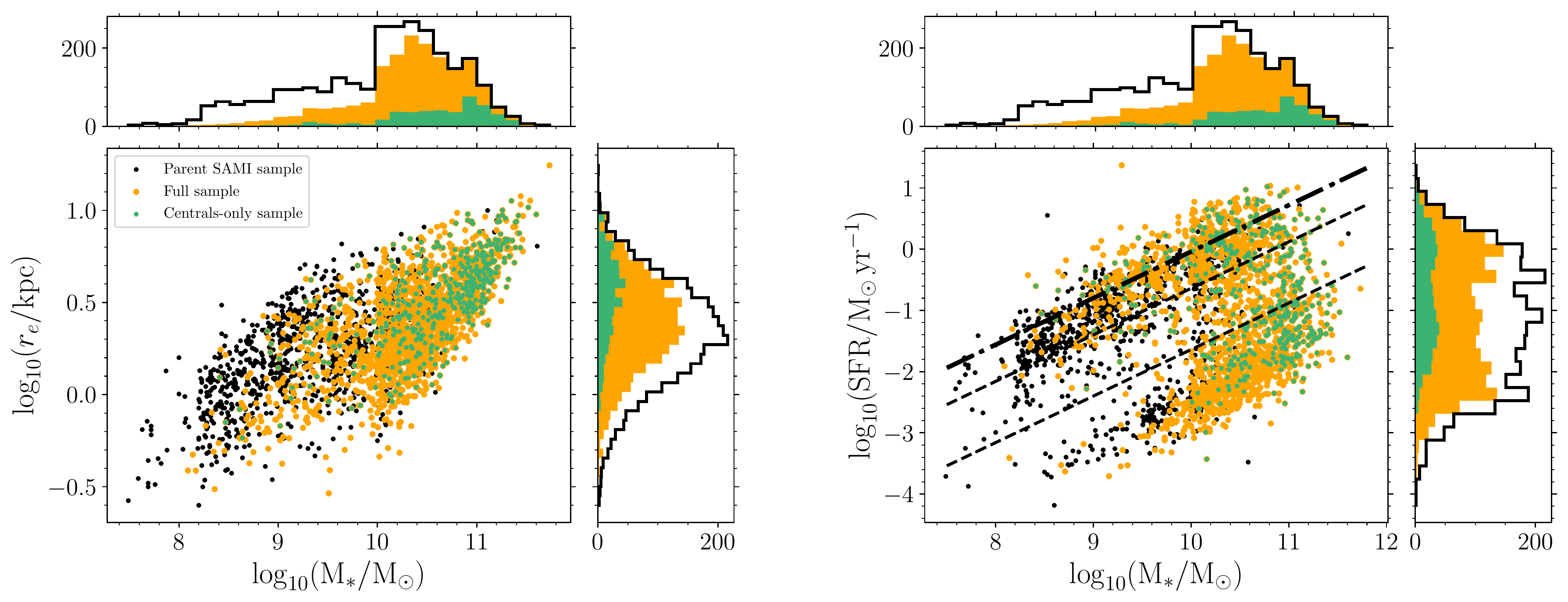}
    \caption{The mass-size plane (left) and mass-SFR plane(right) of our sample. Each panel shows the parent SAMI catalogue (black), the galaxies which form our "full" sample (orange) and those which are part of the "centrals-only" sample (green), as discussed in Section \ref{sec:stellar_pop_measurements}. Histograms show the distribution of each parameter. In the mass-SFR plane, we also show the star forming sequence from \protect\cite{Renzini:2015}, which we use to classify galaxies as star-forming, passive or intermediate: star-forming galaxies are above the highest dashed line; quenched galaxies are below the lower dashed line; intermediate galaxies are between the dashed lines.}
    \label{fig:sample_selection}
\end{figure*}

Finding direct evidence of quenching in an individual object by catching such processes in the act is notoriously difficult, however, since we lack the ability to study individual galaxies across cosmic time. The few examples we do have tend to be of extremely rapid quenching via the removal of a galaxy's supply of cold gas, such as in the case of "Jellyfish" galaxies falling into massive cluster potentials (e.g. \citealt{Gunn:1972}; \citealt{Smith:2010}; numerous observations from the "GAs Stripping Phenomena in galaxies with MUSE" survey of \citealt{Poggianti:2017}). Whilst spectacular, these examples are clearly not typical of the type of quenching which the majority of galaxies in the Universe undergo.

Another route to approach the problem is by studying a large sample of galaxies and investigating the properties of the entire population as a whole. This was the approach taken by \citet[][hereafter \citetalias{Peng:2015}]{Peng:2015} and \citet[][hereafter \citetalias{Trussler:2020}]{Trussler:2020}, who used the stellar ages and metallicities of tens of thousands of galaxies from the Sloan Digital Sky Survey (SDSS; \citealt{York:2000, Eisenstein:2011, Blanton:2017}) in the nearby Universe to conclude that quenching processes must be slow in nature, with quenching timescales of the order of 4 Gyrs.

Key to their conclusions is the correlation between stellar mass and stellar metallicity, and the fact that quiescent galaxies tend to be more metal rich than star forming galaxies at fixed mass \citepalias{Peng:2015, Trussler:2020}. Previous studies have discussed that the size of a galaxy is also an important parameter to consider when analysing its stellar populations \citep[e.g.][]{McDermid:2015, Scott:2017}.  Furthermore, recent work by \citet{Barone:2018}, \citet{D'Eugenio:2018} and \cite{Barone:2020} has quantified this fact: they find that the metallicity of gas and stars in galaxies forms a tighter relationship with a combination of mass and size  ($\Phi = \log_{10}\left(\frac{M_*}{M_{\odot}} \right) - \log_{10}\left(\frac{r_e}{\mathrm{kpc}} \right)$, a proxy for the galaxy's gravitational potential) compared to the relationship between metallicity and stellar mass alone. 

Given that, at fixed stellar mass, star-forming galaxies are larger than passive galaxies \citep{Shen:2003,Lange:2015}, they also have shallower gravitational potentials. Taken together, these results beg the question of whether the difference in metallicity between passive and star-forming galaxies at fixed stellar mass discussed by \citetalias{Peng:2015} and \citetalias{Trussler:2020} is still evident at fixed $\Phi$. Can quenching processes which depend on both mass \textit{and size} be used to explain the diversity of galaxies we observe today in the local Universe?

In this work, we build on the previous studies of \citet{Barone:2018,Barone:2020} and \citet{D'Eugenio:2018} and, for the first time, investigate the correlations between stellar metallicity, stellar mass and gravitational potential for a homogeneously-observed sample containing both quiescent and star-forming galaxies together, instead of studying each class of galaxy separately. We then ask a similar question to \citetalias{Peng:2015} and \citetalias{Trussler:2020}: what can we learn about the quenching timescales of galaxies from these observations? One key difference between our sample and that of \citetalias{Peng:2015} and \citetalias{Trussler:2020} is that each galaxy we study has a robust half-light radius measurement, allowing us to investigate the roles of mass and size in regulating galaxy quenching processes and the production of metals. We also have the advantage of using spatially resolved observations instead of the fixed apertures used by \citetalias{Peng:2015} and \citetalias{Trussler:2020}. This means that we can limit the common issue of aperture bias resulting from probing galaxies of different apparent sizes (and radial gradients in metallicity) to varying degrees. 

The outline of this paper is as follows. In Section \ref{sec:sample_selection} we briefly introduce the SAMI galaxy survey and describe the stellar mass and half-light radius measurements we will be using. In Section \ref{sec:stellar_pop_measurements} we describe our sample of galaxies and outline our method to measure stellar metallicities. Section \ref{sec:models} describes our techniques for fitting straight lines to our measurements to infer metallicity gradients and central values. In Section \ref{sec:results_section} we present our results and in Section \ref{sec:discussion} we discuss our findings, introducing a simple toy model of galaxy quenching to reproduce the population of local galaxies in Section \ref{sec:toy_models}. In Section \ref{sec:conclusions} we draw our conclusions. We assume a flat $\Lambda$ cold dark matter ($\Lambda$CDM) cosmology with $\Omega_{\Lambda}=0.7$, $\Omega_M = 0.3$, and $H_0 = 70$kms\textsuperscript{$-1$}Mpc\textsuperscript{$-1$}. We also assume a \cite{Chabrier:2003} initial mass function where necessary.

\section{The SAMI Survey and Ancillary Data}
\label{sec:sample_selection}

The galaxies in this work are drawn from data release 3 (DR3) of the SAMI galaxy survey \citep{Croom:2021}. SAMI DR3 provides fully reduced datacubes for 3068 galaxies in the local Universe, selected to have a redshift ($z$) less than 0.11 and a stellar mass between $7.8 \leq \log_{10}(\mathrm{M}_{*} / \mathrm{M}_{\odot}) \leq 11.8$. The SAMI instrument \citep{Croom:2012} is mounted at the prime focus on the Anglo-Australian Telescope that provides a 1 degree diameter field of view. SAMI uses 13 fused fibre bundles ("Hexabundles"; \citealt{Bland-Hawthorn:2011, Bryant:2014}) with a high (75\%) fill factor. Each bundle contains 61 fibres of 1.6\arcsec diameter resulting in each IFU having a diameter of 15\arcsec. The IFUs, as well as 26 sky fibres, are plugged into pre-drilled plates using magnetic connectors. SAMI fibres are fed to the double-beam AAOmega spectrograph \citep{Sharp:2006}. AAOmega allows a range of different resolutions and wavelength ranges. The SAMI Galaxy survey uses the 570V grating at 3700-5700\AA{} giving a resolution of $R=1730$ ($\sigma=70.4$ km s$^{-1}$), and the R1000 grating from 6250-7350\AA{} giving a resolution of $R=4500$ ($\sigma=29.6$ km s$^{-1}$).

For each galaxy, we also require a number of ancillary measurements for our analysis. Stellar masses for each SAMI galaxy were derived using $g$ -- $i$ colours and the prescription of \cite{Taylor:2011}, updated in \citet{Bryant:2015}. Total star formation rates (SFRs) were derived from H$\alpha$ flux measurements by \citet{Medling:2018}, with an aperture correction applied where appropriate. As discussed in \citet{Medling:2018}, these SFR measurements should be considered as lower limits, since spaxels with contribution to the H$\alpha$ flux from shocks or AGN are removed before a global SFR is calculated. Furthermore, low surface brightness H$\alpha$ emission may be below the detection limits adopted. This will tend to bias low-mass galaxies to lower SFRs. We therefore choose to use the relation between SFR and stellar mass from \citet{Renzini:2015} to classify galaxies as star-forming, intermediate or quiescent, rather than fitting to the SAMI SFR and mass measurements ourselves (see Section \ref{sec:stellar_pop_measurements}).

Half-light radii were measured using the Multi-Gaussian expansion method of \citet{Cappellari:2002} from  $r$-band images of each object \citep{D'Eugenio:2021}. The photometry is drawn from the Sloan Digital Sky Survey data release 7 \citep{Abazajian:2009} and 9 \citep{Ahn:2012}, as well as the VLT Survey Telescope’s ATLAS Survey (\citealt{Shanks:2015}). See \citet{D'Eugenio:2021} for further details. 

\section{Stellar Population Measurements and Sample Selection}
\label{sec:stellar_pop_measurements}

We measure stellar population parameters as a function of radius for each galaxy in the sample. A full description of these measurements used in this work will be presented in a forthcoming paper (Vaughan et al. in prep). We give a brief summary here.  Firstly, the blue and red arm spectra are joined together and convolved with a Gaussian kernel (of variable width) such that the spectral resolution is a constant value at all wavelengths (followong \citealt{vandeeSande:2017}). We then use the Voronoi binning algorithm of \citet{Cappellari:2003} to aggregate spectra together such that their combined spectrum has a minimum signal-to-noise (S/N) ratio of 20 \AA\textsuperscript{-1} in the V-band. Individual spaxels with a S/N above 20 \AA\textsuperscript{-1} are left unbinned. 181 galaxies lacked the required S/N to create a single Voronoi bin and were discarded. At this stage, we remove from our sample any galaxies which have fewer than 10 independent Voronoi bins or have \lmstar $< 8$. We also remove interacting galaxies or galaxies with close companions. This leaves us with 78,531 spectra from 1363 galaxies, which we call the "full sample". 

We then use the penalised-pixel fitting code \textsc{pPXF} \citep{Cappellari:2004,Cappellari:2017} to fit the MILES simple stellar population (SSP) models of \cite{Vazdekis:2015} to the spectrum from each Voronoi bin. The templates range in metallicity from -2.21 dex to 0.4 dex, in age from 30 Myrs to 14 Gyrs and in [$\alpha$/Fe] abundance from 0.0 to +0.4 dex. The templates use the isochrones from the "Bag of Stellar Tracks and Isochrones" models (BaSTI; \citealt{Pietrinferni:2004, Pietrinferni:2006}). 

In Appendix \ref{sec:BPASS}, we investigate the use of the Binary Population and Spectral Synthesis models (BPASS: \citealt{Eldridge:2017, Stanway:2018,Byrne:2022}) as an alternative to the MILES SSPs. We find the stellar metallicity values derived using BPASS are offset to lower values of [Z/H], and there is also a difference in the derived slopes of the correlations between stellar metallicity and mass and stellar metallicity and $\Phi$. However, our overall conclusions from Section \ref{sec:results_section} remain unchanged. See Appendix \ref{sec:BPASS} for further details.

In essence, \textsc{pPXF} finds the weighted sum of SSP templates which best recovers the input galaxy spectrum. We also include gas emission line templates corresponding to the Balmer series (H$\alpha$, H$\beta$ and H$\gamma$) as well as the atomic species [\ion{N}{II}], [\ion{O}{III}], [\ion{S}{II}], and [\ion{O}{I}]. We use a multiplicative Legendre polynomial of order 10 to correct for small differences in the shape of the observed and template spectra. We normalise the SSP spectra  such that the resulting stellar population parameters are mass-weighted. We also measured the stellar population parameters from the spectra extracted from within the half-light radius ($r_e$) of each galaxy in the same manner, and find that our conclusions remain unchanged.

We estimate the uncertainties on our mass-weighted metallicity measurements by selecting a sample of 10,000 spectra and bootstrapping their measurements 50 times. We then build a simple model to predict this bootstrapped uncertainty ($\sigma_{[Z/H]}$) from the spectrum's [Z/H] and signal-to-noise ratio. This model is then applied to the remaining spectra. Further details of our uncertainty modelling are given in Appendix \ref{sec:uncertainties}.

Finally, we also create a "central-only sample" to differentiate between galaxies which are dominant in their dark matter halos ("centrals") and those which are not ("satellites"). We use the Galaxy and Mass Assembly (GAMA) galaxy group catalogue (G\textsuperscript{3}C) of \cite{Robotham:2011}. In practice, the G\textsuperscript{3}C catalogue selects objects which are the brightest galaxy in the $r$-band within each group to be "centrals". It should be noted that these galaxies are not guaranteed to be the most massive in their group. We include isolated galaxies as centrals. The goal of the central-only sample is to create a sample of objects which have primarily undergone (or will undergo) \textit{internal} quenching processes (in contrast to quenching processes which depend on environment, which predominantly influence satellite galaxies: see e.g. \citealt{Woo:2017}). We will use this sample to compare to the quenching toy model presented in \ref{sec:toy_models}, since the model does not include any prescription for environmental quenching.  Our central-only sample contains 527 galaxies, 39\% of the full sample. 

\begin{figure*}
    \centering
    \includegraphics[width=\textwidth]{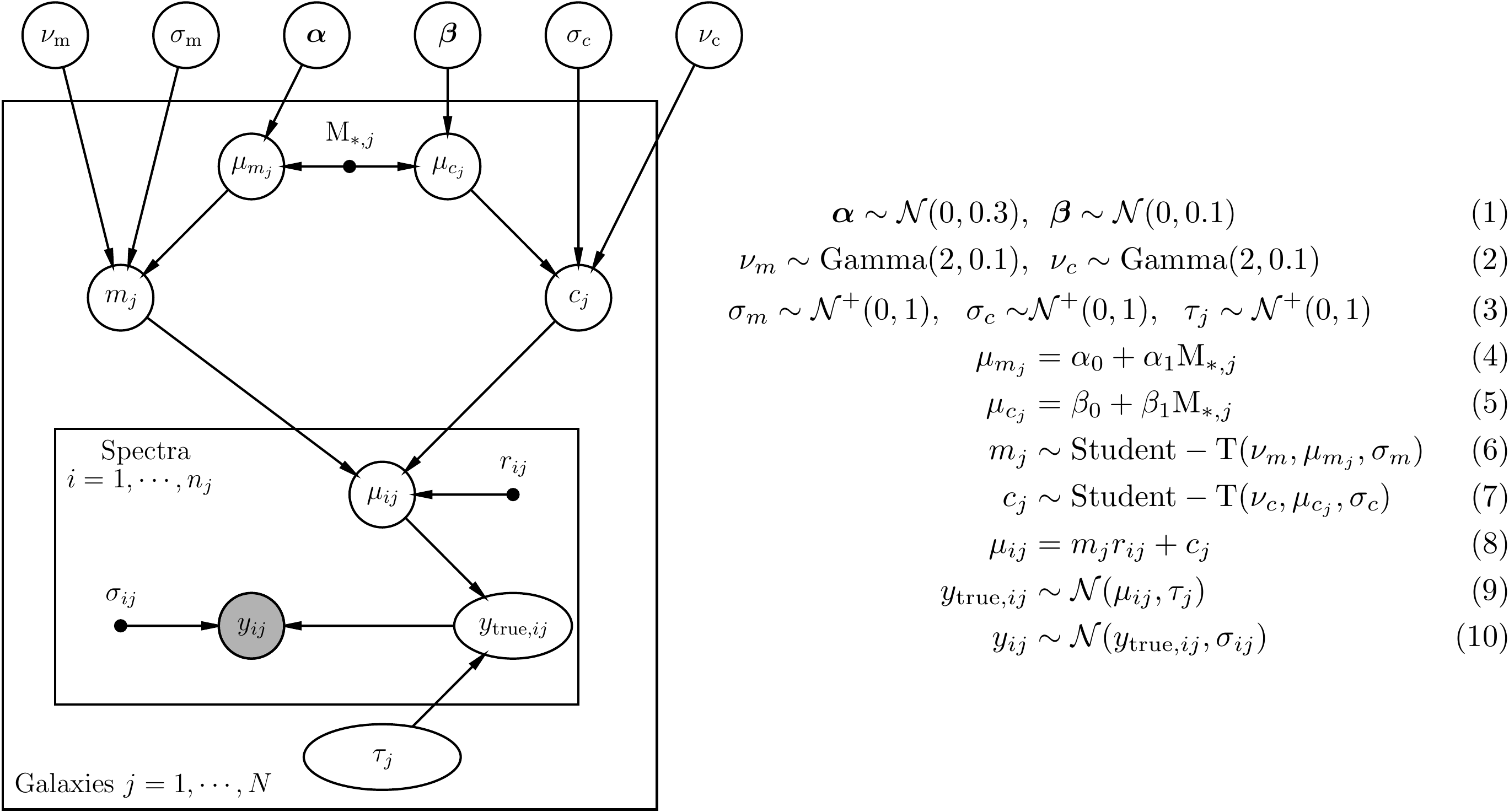}
    \caption{A graphical representation of our hierarchical model (left) and a mathematical description of the priors and hyperpriors used (right). See Section \ref{sec:more_complicated} for a full discussion.}
    \label{fig:hierarchical_model_graphical_representation}
\end{figure*}

Plots of the mass-size and mass-SFR planes for the SAMI parent catalogue and our samples are shown in Figure \ref{fig:sample_selection}. We classify galaxies as being star-forming or passive using simple cuts in the mass-SFR plane. Following \cite{Renzini:2015}, we define the star-forming "main sequence" to be

\begin{equation}
\log(\mathrm{SFR}) = 0.76 \log(\mathrm{M}_{*}/\mathrm{M_{\odot}}) - 7.64
\end{equation}

\noindent with a scatter around the relation of 0.3 dex. We define the difference between a galaxy's measured star-formation rate (SFR) and the SFR of a galaxy of the same mass on the main sequence as $\Delta \mathrm{SFR}_{\mathrm{MS}}$. Following other papers in the literature (e.g. \citealt{Croom:2021a}), we separate galaxies into star-forming, intermediate (or "green-valley") or passive in the following manner:

\begin{align}
\mathrm{SF\,galaxies}&:\Delta \mathrm{SFR}_{\mathrm{MS}}> -0.6\\
\mathrm{Green-valley\,galaxies}&: -0.6 \leq \Delta \mathrm{SFR}_{\mathrm{MS}} \leq -1.6\\
\mathrm{Passive\,galaxies}&:  \Delta \mathrm{SFR}_{\mathrm{MS}} < -1.6
\end{align}

In essence, this definition requires star-forming galaxies to be within 2$\sigma$ (0.6 dex) of the main sequence; passive galaxies to be more than 5$\sigma$ from the main sequence (1.6 dex); and intermediate galaxies to fall between these two categories. Reasonable changes to these values have no effect on our conclusions. Our full sample consists of 442 star-forming galaxies, 694 passive galaxies and 227 intermediate galaxies.

\section{Measuring metallicity gradients and intercepts}
\label{sec:models}

For every galaxy in the full sample, we infer the stellar metallicity at the very centre of the galaxy (galacto-centric radius $r=0$) by fitting a straight line to the metallicity measurements as a function of radius. We perform this fit in two different ways, with our conclusions the same in each case. A reader uninterested in the technical details of the fitting can safely skip this section and move on to the results in Section \ref{sec:results_section}. We note that our conclusions are also unchanged if we instead use the stellar metallicities found from the 1$r_e$ spectra instead of the central metallicity of each galaxy, as shown in Appendix \ref{sec:central_vs_1re_comparison}.

\subsection{Simple straight line fits}
\label{sec:simple_model}

First, we perform a simple straight-line fit to the [Z/H] measurements in each galaxy as a function of $r/r_{e}$, where $r_e$ is the half-light radius of the galaxy. For reasons which will become clear shortly, we refer to this as the "unpooled" fit. 

For each galaxy $j$ we measure a metallicity gradient, $m_j$, metallicity at $r=0$, $c_j$, and the intrinsic scatter around the best-fit straight line, $\tau_j$. If each galaxy has $n_j$ metallicity measurements, $y_{ij}$, at radius $r_{ij}$ with uncertainty $\sigma_{ij}$, our likelihood function for each galaxy takes the form

\begin{equation}
\label{eqtn:likelihood_function}
\log \mathcal{L}_{j} \propto \sum_{i=0}^{i=n_{j}} \left(\frac{(y_{ij} - (m_j r_{ij} + c_j))^2}{\sigma_{ij}^2 + \tau_j^2 }\right)
\end{equation}

To perform the fit, we apply weakly informative Gaussian priors of $\mathcal{N}(0, 1)$ to each $m_j$ and $c_j$, and weakly informative half-Gaussian priors of $\mathcal{N}^{+}(0, 1)$ to each $\tau_j$ (to ensure the $\tau_j$ always remain positive).\footnote{$\mathcal{N}(\mu, \sigma)$ refers to a Gaussian distribution of mean $\mu$ and standard deviation $\sigma$. $\mathcal{N}^{+}(\mu, \sigma)$ is a "half-Gaussian" function which is constrained to only be non-zero for positive values of the dependent variable.} 

We perform the fit using the probabilistic programming language \textsc{Stan} \citep{stan} and its python interface \textsc{pystan}\footnote{\url{https://pystan.readthedocs.io/en/latest/index.html}}, which performs full Bayesian inference of the posterior distribution using an implementation of the No U Turn sampler (NUTS; \citealt{Hoffman:2014}). We performed 1000 warm-up transitions and 1000 draws from the posterior using 4 independent chains.  We ensured that there were no divergent transitions and that the  potential scale reduction factor $\hat{R}$ \citep{Gelman:1992} was always within acceptable limits (between 1 and 1.1). We take the best fitting value of each parameter to be its posterior mean and its uncertainty to be the standard deviation of the posterior samples.

Our unpooled fit suffers from two issues. Firstly, galaxies with few metallicity measurements will have large uncertainties on their derived parameters. Secondly, the highest and lowest metallicity templates used in this work are +0.4 dex and -2.21 dex, and 695 and 78 spectra hit these upper and lower limits respectively. The correct way to treat these measurements is as "right censored" or "left censored" data, since we only know that their true metallicity value is \textit{at least} 0.4 dex or \textit{at most} -2.2 dex. 

We now build a model to address each of these issues. 

\subsection{More complicated fits: hierarchical modelling and censored data}
\label{sec:more_complicated}

We also measure $m_j$, $c_j$ and $\tau_j$ for each galaxy using a hierarchical Bayesian model. The idea of a hierarchical model is to share or "pool" information between "groups" within a set of data, in order to obtain tighter constrains than in the unpooled case above. Here, our metallicity measurements are naturally grouped by the galaxy they are observed from. For galaxies with few metallicity measurements, where the unpooled model may return more uncertain results, a hierarchical model can "share" constraints from similar galaxies to get the tightest inferences on $m$ and $c$.

Another way to think about the hierarchical model is in terms of the priors we apply to each galaxy. In the unpooled model, these priors are identical and fixed for each object. In a hierarchical model, we use hyperparameters to modify the priors for each galaxy depending on other quantities of interest (in this case, stellar mass). This allows us to place different priors on the central metallicity a galaxy of $10^{11}$\Msun{} compared to a galaxy of $10^{8}$\Msun{} (for example). The key point is that we do not put these different priors in by hand; \textit{they are also estimated from the data during the fitting process}. An introduction to hierarchical models can be found in \cite{BDA3}, and some examples of their use in astronomy can be found in \cite{Lieu:2017, Varidel:2020, Grumitt:2020} and \cite{Vaughan:2020}.

\begin{figure*}
    \centering
    \includegraphics[width=\textwidth]{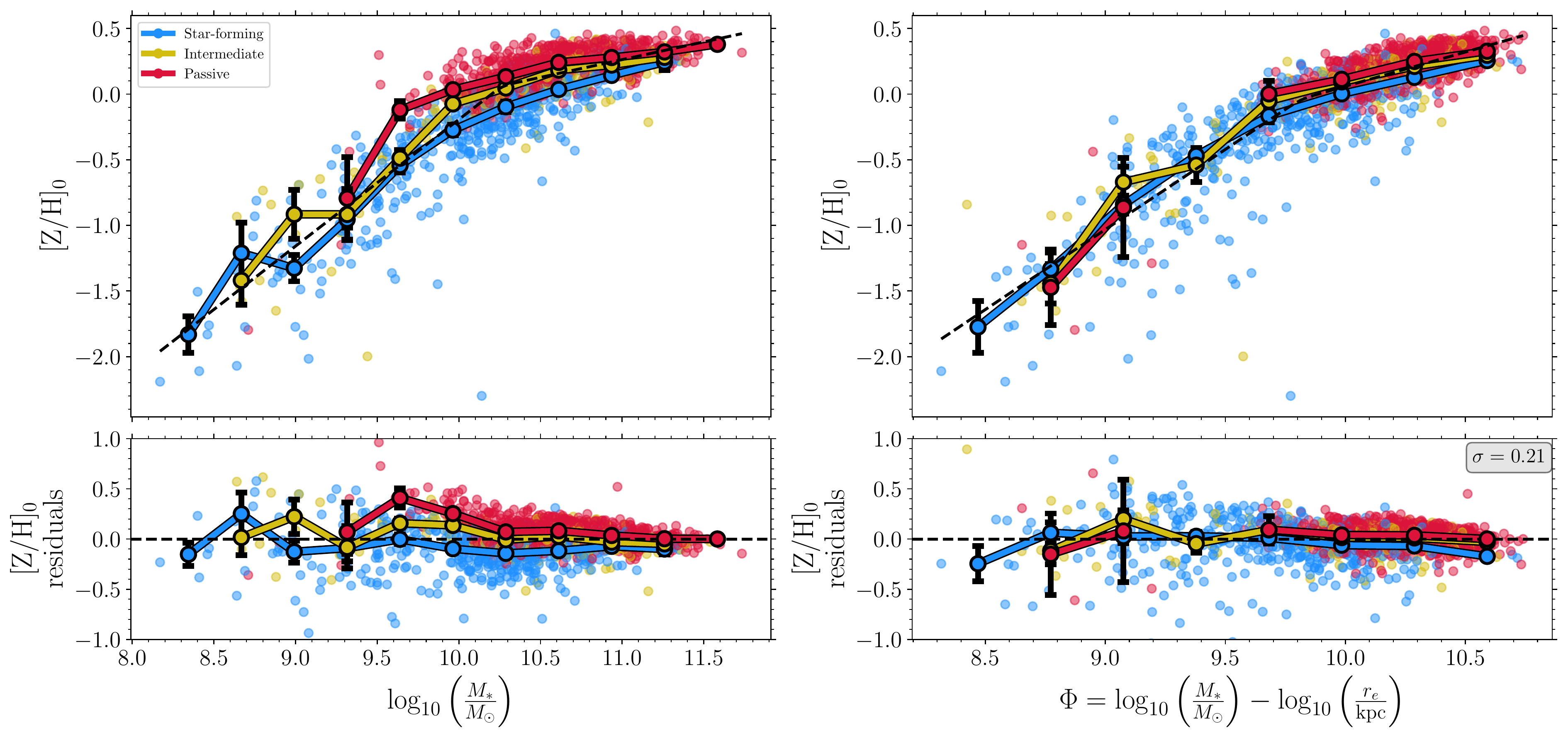}
    \caption{Mass-metallicity (left) and $\Phi$-metallicity (right) relations for our SAMI galaxies. The metallicity measurements are mass-weighted and measured at the centre of each galaxy (see Section \ref{sec:stellar_pop_measurements}). We separate our sample into passive, star-forming and intermediate ("green valley") galaxies according to the definition in \protect\cite{Renzini:2015}. The large circles and coloured lines show the median value of [Z/H]$_{0}$ for star-forming, intermediate and passive galaxies in bins of width 0.3 dex. Black error bars show the standard error on the median. The dashed lines in each panel show a broken-line fit to all galaxies. The lower panels show the residuals around the dashed lines as a function of stellar mass (left) and gravitational potential (right). We measure the intrinsic scatter around the [Z/H]$_{0}$-$\Phi$ relation to be 0.21 dex. The offsets between passive, intermediate and star-forming galaxies in the mass-metallicity plane are in very good agreement to those measured in \protect\cite{Trussler:2020} and \protect\cite{Peng:2015}. The right-hand panel shows that all galaxies form a much tighter sequence in the [Z/H]$_{0}$-$\Phi$ plane, showing that accounting for galaxy size reduces the difference between star-forming and passive galaxies.}
    \label{fig:mass_potential_metallicity_passive_SF}
\end{figure*}

Figure \ref{fig:hierarchical_model_graphical_representation} shows a graphical representation of our hierarchical model. We place a Student's T distribution\footnote{A Student's T distribution is described by three parameters; a location ($\mu$), a scale ($\sigma$) and a "degree of freedom" or "normality"
 parameter ($\nu$). As $\nu$ becomes large, the Student's T probability distribution tends towards a Gaussian.} prior on the gradients, $m_j$, and intercepts, $c_j$ (Figure \ref{fig:hierarchical_model_graphical_representation}, lines 6 and 7). The central location of these priors, $\mu_{mj}$ and $\mu_{cj}$, vary linearly with stellar mass (Figure \ref{fig:hierarchical_model_graphical_representation} lines 4 and 5) with coefficients $\bm{\alpha}=(\alpha_0, \alpha_1)$ and $\bm{\beta}=(\beta_0, \beta_1)$ respectively. These priors incorporate the information that  metallicity, metallicity gradients and galaxy stellar mass are correlated and can be  described by linear relations (e.g. \citealt{Gallazzi:2005}, \citetalias{Peng:2015, Trussler:2020}, \citealt{Goddard:2017a, Zheng:2017, Li:2018, Santucci:2020}). The remaining hyperparameters are given fixed priors as described in Figure \ref{fig:hierarchical_model_graphical_representation} (lines 1 through 3). 

For the spectra with metallicity values at the limits of our templates, we modify the likelihood function to treat these observations as "censored" data. Using our Gaussian likelihood function, the probability that a measurement has a value greater than a value $U$ is given by

\[
\mathrm{Pr}[y>U] = \int_{U}^{\infty}\mathcal{N}(y|\mu, \sigma)dy
 = 1 - \Phi\left(\frac{y - \mu}{\sigma}\right)
 \]

where $\Phi$ is the standard normal cumulative distribution function. We can therefore simply sum each censored contribution to the log likelihood function as

\[
\log \mathcal{L}_\mathrm{right\,censored} = \sum_{i,j} \left(1 - \Phi\left(\frac{y_{ij} - \mu_j}{\sigma_j}\right)\right)
 \]
 
and add this contribution to the likelihood function for the uncensored data.  Similarly, for data less than a value $L$

\[
\mathrm{Pr}[y<L] = \int_{-\infty}^{L}\mathcal{N}(y|\mu, \sigma)dy
 = \Phi\left(\frac{y - \mu}{\sigma}\right)
 \]
and 
\[
\log \mathcal{L}_\mathrm{left\,censored} = \sum_{i,j}  \Phi\left(\frac{y_{ij} - \mu_j}{\sigma_j}\right)
 \]

We again use \textsc{Stan} to perform perform the fitting, again performing 1000 warm up transitions and 1000 draws from the posterior for 4 independent chains. We take the best fitting value of each parameter to be its posterior mean and the standard deviation of the posterior samples to be the uncertainty.

The results in this paper all use the hierarchical model, although we find that our conclusions are unchanged if we instead use the simple unpooled results. A quantitative comparison of the two is presented in Appendix \ref{sec:hierarchical_normal_comparison}.

\section{Results}
\label{sec:results_section}

The main result of this work is the relationship between the central metallicity of galaxies ([Z/H]$_{0}$, as measured from their radial metallicity gradients)  and a proxy for their gravitational potential, $\Phi = \log_{10}\left(\frac{M_*}{M_{\odot}} \right) - \log_{10}\left(\frac{r_e}{\mathrm{kpc}} \right)$. This is presented in Figure \ref{fig:mass_potential_metallicity_passive_SF}.

We show the results of our central metallicity measurements as a function of mass in the left hand panel of Figure \ref{fig:mass_potential_metallicity_passive_SF}, with SF galaxies coloured blue, passive galaxies shown in red and intermediate galaxies in yellow. Quantitative results of fitting a straight line to the star-forming and passive galaxies are shown in Table \ref{tab:straight_line_fits}. 

\begin{table}
\centering
\caption{Results from the straight line fits to the mass-metallicity relation for star-forming, passive and intermediate ("green valley") galaxies. We fit relations of the form $[\mathrm{Z/H}] = m(\log_{10}(\mathrm{M}_* / \mathrm{M}_{\odot}) - 10.3) + c$. The difference in metallicity between galaxies at a fixed mass of 10$^{10.3}$\,M$_{\odot}$ is in very good agreement with \protect\cite{Peng:2015} and \protect\cite{Trussler:2020}.}
\label{tab:straight_line_fits}
\begin{tabular}{lll}
\toprule
{} &            $m$ &             $c$ \\
\midrule
Star-forming  &  $0.72\pm0.02$ &  $-0.07\pm0.01$ \\
Intermediate &  $0.57\pm0.02$ &  $-0.00\pm0.01$ \\
Passive      &  $0.26\pm0.01$ &   $0.16\pm0.00$ \\
\bottomrule
\end{tabular}
\end{table}

We confirm the well known result that passive and SF galaxies occupy different areas of the mass-metallicity plane, and that passive galaxies tend to be more metal-rich than SF galaxies at the same mass. The straight line fits to the mass-metallicity relation are in excellent agreement with the recent work of \citetalias{Peng:2015} and \citetalias{Trussler:2020}; we find that, at a stellar mass of $\log(\mathrm{M}_{*}/\mathrm{M_{\odot}})=10.3$, the offset in metallicity between star-forming and passive galaxies is 0.23 dex, compared to 0.2--0.25 dex from \citetalias{Trussler:2020}. We also measure a difference in metallicity between star-forming and intermediate galaxies to be 0.07 dex at $\log(\mathrm{M}_{*}/\mathrm{M_{\odot}})=10.3$, very similar to the 0.06--0.012 dex measured in \citetalias{Trussler:2020}.

\begin{figure*}
    \centering
    \includegraphics[width=\textwidth]{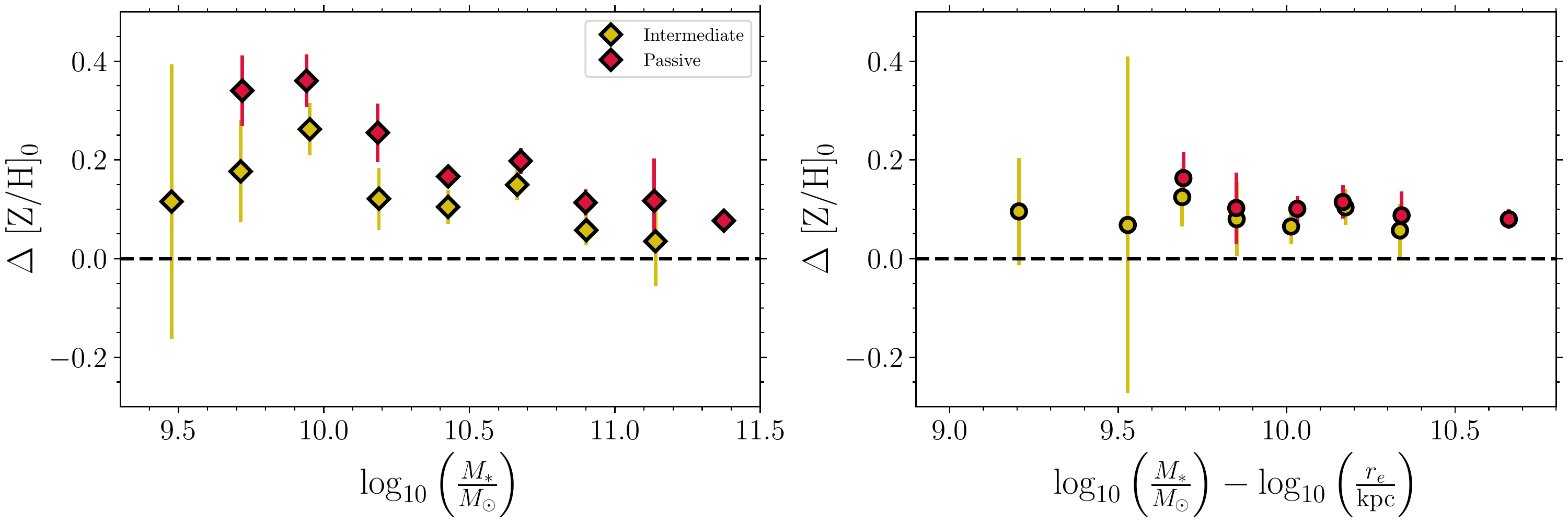}
    \caption{The difference in metallicity between star-forming and passive galaxies (red points) and star-forming and intermediate galaxies (yellow points) as a function of mass (left) and $\Phi$ (right). Each point represents the median metallicity difference of our sample in bins of mass (left) and $\Phi$ (right). We only show bins which contain at least 5 galaxies. Error bars represent the sum in quadrature of the standard error on the median of the star-forming bin and the passive/intermediate bin as appropriate. }
    \label{fig:delta_ZH}
\end{figure*}

This difference in metallicity between star-forming and passive galaxies is mass dependent, dropping to a smaller value of 0.07 dex at the high-mass end of our sample, again in agreement with \citetalias{Peng:2015} and \citetalias{Trussler:2020}. This is shown in the left hand panel of Figure \ref{fig:delta_ZH}.

The right hand panel of Figure \ref{fig:mass_potential_metallicity_passive_SF}, shows the mass-weighted central metallicity of each galaxy against a proxy for its gravitational potential, $\Phi$, where $\Phi = \log_{10}\left(\frac{M_*}{M_{\odot}} \right) - \log_{10}\left(\frac{r_e}{\mathrm{kpc}} \right)$.  This plot shows that the metallicity difference between the average star-forming and passive galaxy at fixed $\Phi$, rather than fixed mass, is much smaller. As discussed in \cite{Barone:2018, Barone:2020}, for both passive and star-forming galaxies, stellar metallicity correlates better with $\Phi$ than mass (as quantified by the intrinsic scatter around the relation). Here, we show for the first time that the distributions of star-forming and passive galaxies in the [Z/H]-$\Phi$ plane smoothly follow on from one another. 

We also show the difference in metallicity at fixed $\Phi$ between star-forming, passive and intermediate galaxies in the right-hand panel of Figure \ref{fig:delta_ZH}. It is clear that the metallicity offset at fixed $\Phi$ between passive and star-forming galaxies is reduced compared to when using mass: the largest difference is 0.14 dex at $\Phi$ = 9.5, compared to a difference of 0.38 dex at \lmstar = 9.6, and the average offset across all bins in $\Phi$ is 0.11 dex, compared to the average across all bins in mass of 0.21 dex. It is only for the highest mass bin where the $\Delta$[Z/H] values are comparable, being 0.08 dex for the rightmost points in each panel. The difference at fixed $\Phi$ is also approximately constant, being consistent with a single offset of $0.11\pm0.03$ dex at all values of $\Phi$.

If we use the spectra extracted from 1$r_e$ for each galaxy instead of the central values of [Z/H] inferred from our hierarchical model, we find a very similar result: the average offset in [Z/H] between passive and star-forming galaxies across all bins in $\Phi$ is $0.07\pm0.03$ dex.

We fit a broken line model to all galaxies in the $\Phi$-[Z/H] and mass-[Z/H] planes, described by

\begin{equation}
\label{eqtn:phi_broken_line}
    [\mathrm{Z/H}] {} = 
    \begin{cases}
      c + m_{u}\Phi &\text{if           $\Phi \geq \Phi_b$}\\
      c + m_{l}\Phi  + ( m_{u} - m_l)\Phi & \text{if $\Phi < \Phi_b$}
    \end{cases}
\end{equation}

This broken line has 5 free parameters: the breakpoint, $\Phi_b$; the slope of the line below the break, $m_l$; the slope of the line above the break, $m_u$; an intercept, $c$; and the intrinsic scatter around the relation, $\sigma$. We show the best-fit broken line model for all galaxies as a dotted line in Figure \ref{fig:mass_potential_metallicity_passive_SF}.

We also highlight here how the right-hand panel of Figure \ref{fig:mass_potential_metallicity_passive_SF} might change were we to use half-mass radii rather than half-light radii to calculate $\Phi$, although a full analysis of this point is beyond the scope of this work. As discussed in \cite{Suess:2019}, the half-mass radii of low-redshift quiescent galaxies tend to be a fixed fraction of their half-light radii whereas the ratio of $r_{e, \mathrm{mass}}/r_{e,\mathrm{light}}$ for star-forming galaxies decreases as their mass increases, becoming only $\approx0.6$ for galaxies with \lmstar$\approx11$. This would tend to move high-mass star-forming galaxies to the right in Figure \ref{fig:mass_potential_metallicity_passive_SF},  increasing the scatter around the [Z/H]-$\Phi$ relation and accentuating its break.

\section{Discussion}
\label{sec:discussion}

Many previous studies have discussed the correlations between a galaxy's stellar metallicity and its structural parameters, although predominantly in terms of velocity dispersion $\sigma$ \citep[e.g.][]{Trager:2000, Thomas:2005, Nelan:2005, Graves:2009a, Graves:2009b, McDermid:2015, Scott:2017, Li:2018} rather than $\Phi$ as in this work.

The two quantities are obviously closely related; assuming galaxies are virialised, $ \sigma^2 \propto M/R$. We show this relationship between central velocity dispersion, $\sigma$, and $\Phi = \log_{10}\left( \frac{M_{*}}{M_{\odot}} \right) - \log_{10}\left( \frac{r_e}{\mathrm{kpc}} \right)$ for the full sample in Appendix \ref{sec:sigma_vs_phi}, which clearly shows the expected linear correlation with a gradient of 2 for galaxies with $\sigma$ greater than the SAMI blue-arm instrumental resolution (of $\approx70$ kms$^{-1}$). We use $\Phi$ in this work to allow us to extend Figure \ref{fig:mass_potential_metallicity_passive_SF} to the lowest values of $\Phi$ in our sample, without the complication of measuring velocity dispersion values close to (or below) the instrumental resolution. However,  we note that there may be other, physical reasons why galaxies with low values of $\Phi$ depart from the simple linear relationship with $\sigma$: e.g. perhaps due to an excess of dark-matter, as proposed in \citet{Barat:2019} and \citet{Barat:2020}.

\citet{Barone:2018} and \citet{Barone:2020} were the first to show quantitatively that stellar metallicities correlated tightly with the purely photometric quantity $\Phi\sim \mathrm{M}/r_e$, and that this correlation was tighter than for stellar mass alone. Their work concentrated first on early-type galaxies and then on star-forming galaxies, however, without drawing conclusions about a combined sample. Here, we show for the first time that star-forming, intermediate and quiescent galaxies all lie on the same relation in the [Z/H]-$\Phi$ plane.

The physical mechanism which drives this relation is not yet fully understood, but is likely due to the relationship between mass, size and the local escape velocity of the system \citep[e.g.][]{Franx:1990, Emsellem:1996, Scott:2009, Scott:2017}. Galaxies with deeper potential wells are more able to retain the metal-rich winds and ejecta from massive stars which pollute their interstellar media, meaning that their future generations of stars will be more metal rich than those in galaxies with shallower potential wells. We note however, that this simple pictures complicates if the [Z/H]--Phi relation changes with redshift- a matter we discuss further in Section \ref{sec:validity_of_constant_Z_Phi_relation}.

The fact that galaxies in our sample follow a single [Z/H]-$\Phi$ relation has interesting implications for large photometric surveys of galaxies. Section \ref{sec:results_section} shows that stellar metallicity can be determined solely from photometric observations of $r_e$ and M$_*$ to a 1-$\sigma$ accuracy of $\approx$ 0.2 dex.

To date, estimating stellar metallicities without spectroscopic information, using broad-band colours or via panchomratic spectral energy distribution (SED) fitting, is notoriously unreliable. Many SED fitting codes therefore treat stellar metallicity as a nuisance parameter or fix it at solar metallicity (e.g. \citealt{Skelton:2014}, and see the discussion in \citealt{Conroy:2013}).  Other studies have shown that the choice of different stellar population synthesis models gives different results for the same input colours \citep{Lee:2007, Eminian:2008, Conroy:2013}. Stellar masses, on the other hand, are much more straightforward and robust outputs of SED fitting (due to the fortuitous degeneracies which exist between dust, age, and metallicity: \citealt{Bell:2001, Taylor:2011, vandeSande:2015}).  

Further work into the [Z/H]-$\Phi$ relation could therefore allow for the estimation of stellar metallicities for large catalogues of galaxies without expensive spectroscopic observations for the first time. Whilst the correlation in Figure \ref{fig:mass_potential_metallicity_passive_SF} has till now only been investigated in the low-redshift Universe, calibrating this relationship at higher redshift could allow for studies of stellar metallicity over cosmic time (see also \citealt{Barone:2021}, who find the [Z/H]-$\Phi$ relation of passive galaxies at z=0.7).

\subsection{Implications for galaxy quenching}

\citetalias{Trussler:2020} and \citetalias{Peng:2015} use the difference in metallicity between star-forming and passive galaxies at $z=0$ to conclude that most galaxies in the Universe undergo slow quenching ("starvation"). Their argument is as follows. During the course of stellar evolution, the stars in a galaxy will convert hydrogen and helium into heavier elements. A fraction of the stars in a galaxy will return their metal content to the interstellar medium (through supernovae or mass-loss from stellar winds), which increases the galaxy's overall gas-phase metallicity. On the other hand, accretion of pristine gas from the galaxy's halo or cosmic filaments will tend to dilute a galaxy's gas reservoir and lower its metallicity. The balance of these two processes sets whether a galaxy's next generation of stars will tend to have a lower, higher or similar stellar metallicity to previous generations. 

If a galaxy's supply of cold gas is quickly removed or star formation is quickly suppressed, both of these processes are stopped and the stellar metallicity of a galaxy after quenching is the same as when quenching started. However, if star formation in the disc of a galaxy is allowed to continue but accretion of low-metallicity gas is cut off- i.e. the quenching is slow- this balance is upset and the galaxy's gas-phase metallicity is no longer being diluted. Generations of stars formed during slow quenching will be of a higher metallicity than before, and this process continues until the galaxy's gas is exhausted. Under this scenario, the galaxy's stellar metallicity is much higher after quenching than it was before.

\citetalias{Trussler:2020} and \citetalias{Peng:2015} argue that galaxies evolve along the "main sequence" in the SFR-M$_{*}$ plane and the star forming mass-metallicity relation until slow quenching begins. During quenching, their SFR decreases and their metallicity increases such that they eventually lie in the passive regions of the main sequence and the mass-metallicity relations. This evolution leads to vertical or nearly vertical tracks in the mass-metallicity plane, as a large increase in metallicity is accompanied by only a small increase in stellar mass \citepalias{Peng:2015,Trussler:2020}.

Our findings raise questions about this interpretation. Firstly, Figure \ref{fig:mass_potential_metallicity_passive_SF} shows the importance of having robust size measurements for each galaxy. The near vertical motion in the mass-metallicity plane proposed by \citetalias{Peng:2015} and \citetalias{Trussler:2020} must \textit{also} be accompanied by a decrease in size for the galaxy undergoing quenching to remain on the [Z/H]-$\Phi$ plot from Figure \ref{fig:mass_potential_metallicity_passive_SF} (as long as the assumption of a constant relation  between [Z/H] and $\Phi$ is valid: see Section \ref{sec:validity_of_constant_Z_Phi_relation}).

For an increase in metallicity of 0.2 dex and no change in stellar mass, the right panel of Figure \ref{fig:mass_potential_metallicity_passive_SF} implies that the value of  $\log_{10}\left( \frac{M_{*}}{M_{\odot}} \right) - \log_{10}\left( \frac{r_e}{\mathrm{kpc}} \right)$of the galaxy must change by $\approx$0.4 dex. For no change in stellar mass, the half-light radius of a galaxy must therefore \textit{decrease} by 0.4 dex (i.e. end up at $\sim$40\% of its value before quenching began). Such a reduction in size is not found in simulations of individual galaxies undergoing slow quenching \citep[e.g.][]{Genel:2018}, nor in the recent study of \citealt{Croom:2021a} who investigated the effects of disk-fading on star-forming galaxies and found a reduction in $r_e$ of $\approx10\%$ (0.05 dex). Moreover, a number of studies have found that recently quenched galaxies tend to have large radii (i.e. radii similar to star-forming galaxies) when they join the passive population (e.g. \citealt{Cassata:2013, Carollo:2013, Krogager:2014, Cortese:2019}).

\begin{figure}
    \centering
    \includegraphics[width=\columnwidth]{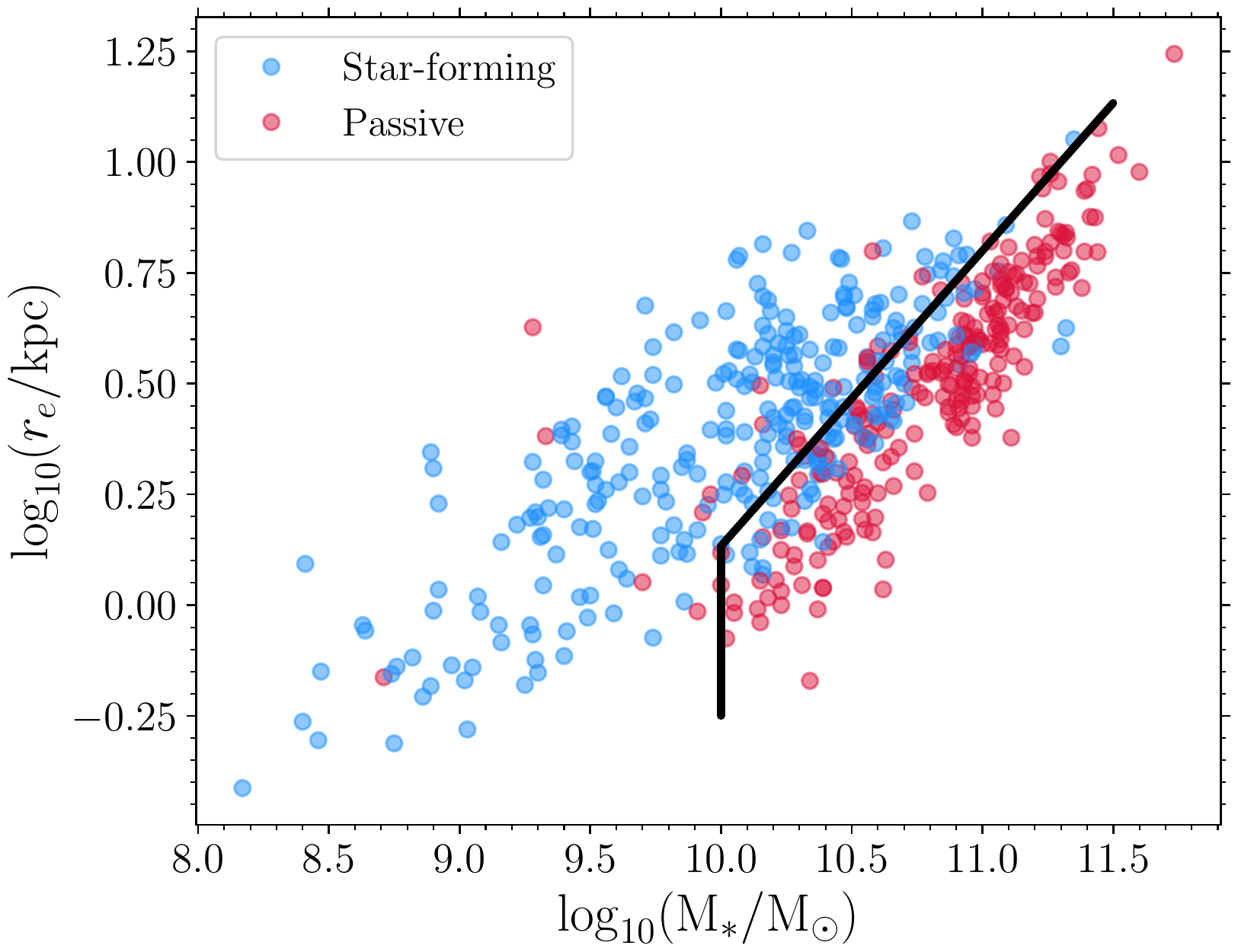}
    \caption{The mass-size plane of the central and isolated galaxies in our sample. Star-forming galaxies are shown in blue and quenched galaxies are shown in red. When a galaxy in our toy model  crosses the thick black line as it grows in mass and size, we assign it a 5\% probability of quenching during the next time step (see Section \ref{sec:toy_models} for details).}
    \label{fig:quenching_model_probability_mass_size_plane}
\end{figure}

\begin{figure*}
    \centering
    \includegraphics[width=\textwidth]{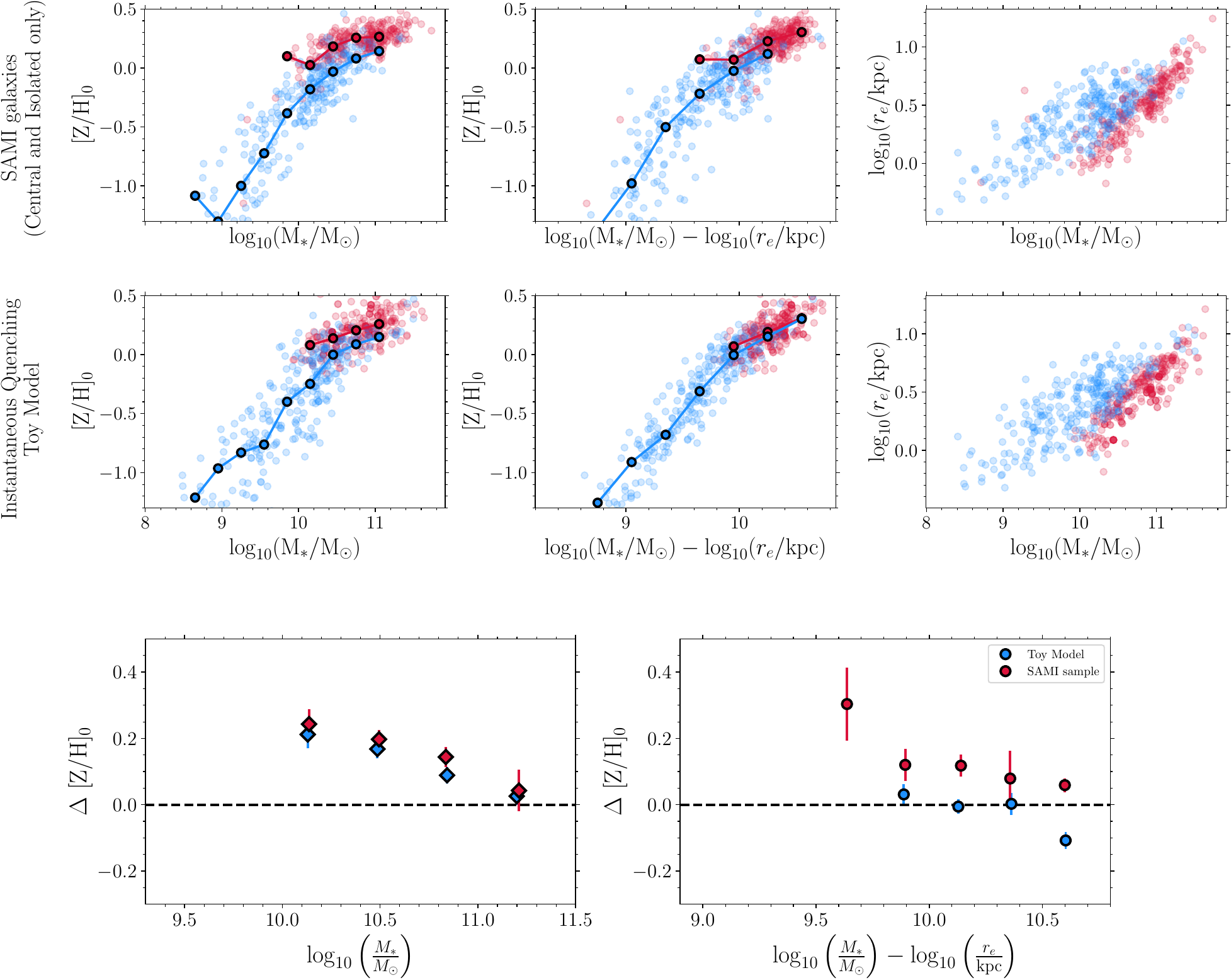}
    \caption{A comparison between the data from our sample of SAMI central and isolated galaxies (top row) to simulated galaxies from our toy model of instantaneous quenching (middle row). The columns show the [Z-H]-M$_{*}$ relation (left), the [Z-H]-$\Phi$ relation (middle) and the mass-size relation (right). The bottom row shows the difference in metallicity between passive and star-forming galaxies as a function of mass (left) and $\Phi$ (right), for the SAMI sample (red points) and the galaxies from the toy model (blue points). We find a very good quantitative agreement between the two samples; in particular, we recover the mass-dependent offset in [Z/H] between passive and star-forming galaxies in our model with instantaneous quenching. We note that the agreement between the model and data in the mass-size panel is by construction; we have sampled our model galaxies to have the same mass and size functions as the SAMI data}. Overall, this model implies that the difference in metallicity between star-forming and passive galaxies at fixed mass cannot, in isolation, be used as evidence of slow quenching.
    \label{fig:quenching_toy_model}
\end{figure*}

\subsection{Quenching toy models}
\label{sec:toy_models}

We now investigate whether we can reproduce our observations using toy models of galaxy quenching. The aim of this Section is to investigate whether a model with almost instantaneous quenching- rather than the slow quenching advocated by \citetalias{Peng:2015} and \citetalias{Trussler:2020}- can simultaneously account for the distribution of galaxies in the mass-metallicity, [Z/H]-$\Phi$ and mass-size planes.

We build a simple model where star-forming galaxies grow in mass and size and evolve in metallicity as they accrete gas and form new stars. These galaxies then have a probability to quench at the start of each timestep, with this probability ($p_{\mathrm{quench}}$) depending on a combination of their mass and their size. Once galaxies become quenched, they stop their evolution and fix their values of mass, size and metallicity until redshift zero. 

We begin by assigning a formation redshift to each galaxy, which is drawn from a uniform distribution between one and six. We next assign galaxies a stellar mass. We use the results from \cite{Davidzon:2017}, who study the galaxy stellar mass function of star-forming and passive galaxies out to redshift six. We use the results from their "active sample" to draw stellar masses from a double Schechter function at $z\leq3$ and a single Schechter function at $z>3$. These represent star-forming galaxies. 

We then place galaxies on the mass-size relation for star-forming galaxies at the redshift of their formation, using measurements from \cite{VdW:2014}. Next, we must assign each galaxy a stellar metallicity. Measurements of the mass-metallicity relation at high redshift are currently limited to studies of ionised gas emission lines. We therefore make a further assumption; that the [Z/H]-$\Phi$ relationship is the same at high-redshift as it is at low redshift (i.e. that measured in Section \ref{sec:results_section}). This assumption is unlikely to be completely correct, and we discuss its validity in Section \ref{sec:validity_of_constant_Z_Phi_relation}.  

We are then left with two free parameters; the intercept of [the Z/H]-$\Phi$ relationship relationship and the intrinsic scatter around it. We choose an intercept of 0.03 dex to match the "combined" relation from Table \ref{tab:straight_line_fits}, and an intrinsic scatter of 0.08 dex. 

Galaxies are evolved forward in time using a timestep of 100 Myr, from their formation redshift to today. To set the amount of stellar mass formed at each timestep, we place star-forming galaxies on the main sequence of star-formation at the appropriate redshift, using the "consensus" relation from \cite{Speagle:2014}. Furthermore, galaxies grow in metallicity according to the "combined" relation described in Section \ref{sec:results_section}. Galaxies increase in size according to:

\begin{equation}
\label{eqtn:dm_dr}
\delta \log R = 0.3 \delta \log M_{*}
\end{equation} 

This evolution in size as a function of mass has been measured by recent observational studies \citep[e.g.][and references therein]{vanDokkum:2013} as well as in simulations \citep{Zolotov:2015}. Other studies find slightly different values (e.g. $\delta \log R\approx0.4\delta \log M_{*}$ in \citealt{Hirschmann:2013}, $\delta \log R\approx0.4\delta \log M_{*}$ in \citealt{Genel:2018}) but small changes to this value do not affect our conclusions.

The final piece is the inclusion of quenching. Figure \ref{fig:quenching_model_probability_mass_size_plane} shows the mass-size plane of the galaxies in our central-only sample. A thick black line shows the region containing the majority of the quenched galaxies in the sample. This region is bounded by:

\begin{enumerate}
    \item \lmstar $ =10$, a line of constant stellar mass.
    \item \lmstar - $\frac{3}{2}\log_{10}(r_e/\mathrm{kpc}) = 9.8$
\end{enumerate}

In the model, if a galaxy grows in mass and size such that it crosses to the right of the line in the mass-size plane, it is assigned a probability of quenching of $p_{\mathrm{quench}}=0.05$ at all future timesteps. Galaxies outside the shaded region have $p_{\mathrm{quench}}=0$. This value of $p_{\mathrm{quench}}=0.05$ was chosen to give a good match between the quenched fraction of the final simulated galaxies and the quenched fraction in the SAMI central-only sample.

At the start of each timestep, we draw a random number on the unit interval and compare with $p_{\mathrm{quench}}$ to decide whether a given galaxy continues to form stars or becomes quenched. If the galaxy quenches at a given timestep, it stops evolving completely in mass, size and metallicity; its values of these quantities at the onset of quenching are frozen in place until redshift zero. This model of galaxy quenching is effectively instantaneous (i.e. completely shutting down star-formation on a timescale of 100 Myrs), unlike the slow quenching (on timescales of Gyrs) discussed in \citetalias{Peng:2015} and \citetalias{Trussler:2020}. We note that our choice of $p_{\mathrm{quench}}=0.05$ per timestep implies that a galaxy in the shaded region of the mass-size plane has a $\approx40$\% chance of quenching after 1 Gyr (10 timesteps).

We create 10$^4$ galaxies and evolve them forwards from their initial redshifts to today. We then compare quantitatively to the central-only SAMI sample discussed in Section \ref{sec:stellar_pop_measurements}, in order to minimise the effects of environmental quenching processes which are not included in our model. This central-only sample contains 527 galaxies. We then randomly sample 527 galaxies from our toy model with the same mass function and size distribution as our SAMI galaxies. Finally, we add reasonable measurement uncertainties to our final model values; a 0.1 dex scatter in stellar mass and a 0.05 dex scatter in [Z/H] and size.

\subsubsection{Results from our model}
\begin{figure*}
    \centering
    \includegraphics[width=\textwidth]{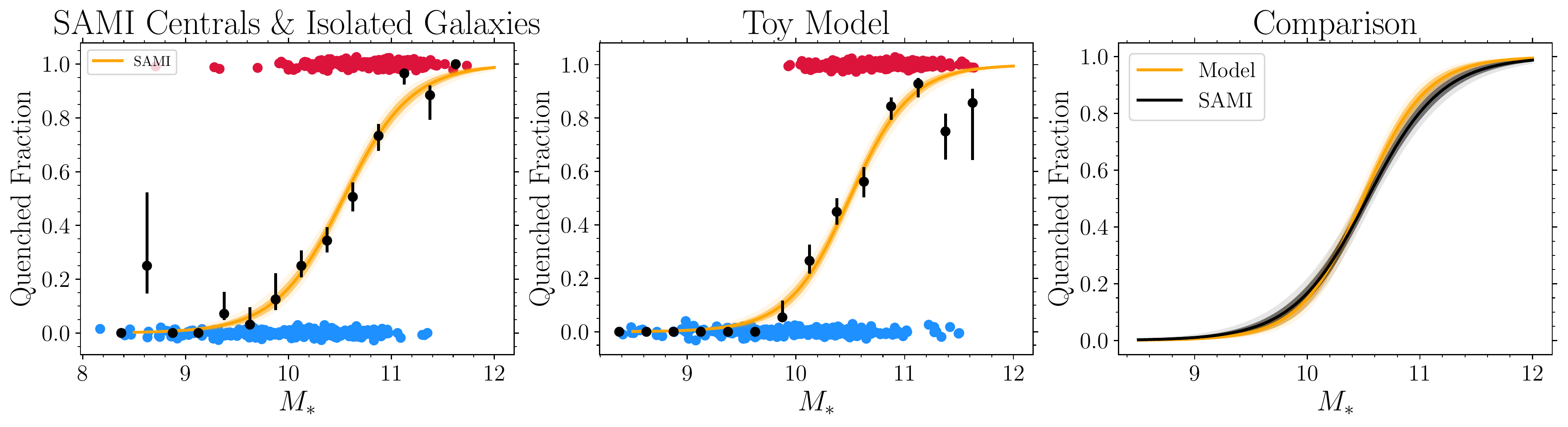}
    \caption{The quenched fraction of galaxies in our our sample of SAMI centrals  (left) and our toy model of quenching (middle). We show whether galaxies are passive (red points, plotted at a $y$ value around 1) or star-forming (blue points, at a $y$ value around 0.0) as a function of their stellar mass. The points have had a small random jitter added such that they don't all lie on top of one another. Black points with error bars show the quenched fraction of galaxies in each bin of stellar mass. The curved lines show logistic regression fits to the same data, with the shaded regions highlighting the 1$-\sigma$ uncertainty. The right panel compares the logistic regression fits for the SAMI centrals (black) and the model galaxies (orange), showing the good agreement between the two.}
    \label{fig:quenching_f_mass_investigation}
\end{figure*}

Our results are shown in Figure \ref{fig:quenching_toy_model}. The top row shows the observed SAMI central and isolated galaxies and the middle row shows results from our instantaneous quenching model. We plot three important relationships: the mass-metallicity relation in the first column; the [Z/H]-$\Phi$ relation in the second panel; and the mass-size relation in the third panel.

\begin{table}
\centering
\caption{
A comparison of the straight-line fits to the mass-metallicity relation for the SAMI centrals and instantaneous quenching toy model shown in Figure \ref{fig:quenching_toy_model}. We fit a relation of the form [Z/H] = $m (\log_{10}(\mathrm{M}_*/\mathrm{M}_{\odot}) - 10.3) + c$. The toy model reproduces the mass-metallicity relation of the SAMI centrals almost exactly. 
}
\label{tab:model_data_comparison}
\begin{tabular}{lllll}
\toprule
{} & $c_{\mathrm{star-forming}}$ & $c_{\mathrm{passive}}$ & $m_{\mathrm{star-forming}}$ & $m_{\mathrm{passive}}$ \\
\midrule
SAMI centrals &              $-0.15\pm0.01$ &          $0.12\pm0.01$ &               $0.61\pm0.03$ &          $0.21\pm0.02$ \\
Toy model     &              $-0.18\pm0.02$ &          $0.09\pm0.01$ &               $0.65\pm0.02$ &          $0.24\pm0.02$ \\
\bottomrule
\end{tabular}
\end{table}

We find that this very simple model can produce an offset in stellar metallicity between star-forming and quiescent galaxies at fixed mass, implying that slow quenching is not the only way to produce this. Our instantaneous quenching model recovers a difference of 0.27 dex between the metallicity of passive and star-forming galaxies at a mass of \lmstar$=10.3$, identical to that from the central-only sample. The existence of this metallicity gap was used by \citetalias{Peng:2015} and \citetalias{Trussler:2020} to conclude that most galaxies undergo slow quenching processes, but this work shows that such a gap can exist even when quenching is nearly instantaneous. In the bottom panel of of Figure \ref{fig:quenching_toy_model}, we also show that we recover the change in this offset as a function of mass in our model. 

We also note that the model does a good job of recovering the overall quenched fraction of galaxies, as well as the quenched fraction of galaxies as a function of mass. In this sample of SAMI centrals, 53\% of galaxies are quenched, compared to the 47\% of galaxies in our toy model which are quiescent. Figure \ref{fig:quenching_f_mass_investigation} shows the distribution of quenched (red points) and star-forming (blue points) galaxies against stellar mass in the model (left panel) and in the SAMI data (middle panel). The black points show the fraction of galaxies that are quenched in mass bins of width 0.25 dex.

We perform logistic regression on the SAMI galaxies and the model galaxies, using their stellar mass as an input and whether they are quenched or star-forming as the binary dependent variable. These are shown by the curved lines in each panel. This is simply another way to compare the quenched fractions as a function of mass between the model and the data. Both the binned fractions and logistic curves show that the model and the data agree well.

In Figure \ref{fig:mass_zh_radius} we show the mass-metallicity relation of our SAMI sample with each point coloured by its effective radius. On the right, we present the same plot for our simulated galaxies. The two panels are in excellent agreement, both showing a clear trend with radius across the relation (see also \citealt{Barone:2018} and \citealt{Barone:2020} who show the same result for early and late type galaxies separately). We note that in all areas of the mass-metallicity plane, at fixed mass galaxies which are smaller have higher metallicity. This is not only true at the high-mass end, where the quenched and star-forming galaxies are offset from one another (as shown in Figure \ref{fig:mass_potential_metallicity_passive_SF}), but for star-forming galaxies at lower masses too. We return to this point in Section \ref{sec:offset_in_MZ_relation} to discuss what we can learn about the reason behind the offset in [Z/H] between quenched and star-forming galaxies.

Finally, we can see from the bottom panel of Figure \ref{fig:quenching_toy_model} that our model underpredicts the offset in [Z/H] as a function of $\Phi$ by a small amount: the average $\Delta$[Z/H] is 0.1 dex for the SAMI sample (0.08 dex if we remove the first outlying point) and -0.02 dex for the toy model galaxies (0.01 dex if we remove the final negative value). This may be pointing to the fact that a process like slow quenching via strangulation, where a galaxy increases its [Z/H] without large changes in \lmstar and $r_e$, is needed. If all galaxies increase their [Z/H] by $\approx$0.05-0.1 dex when they quench, without changing in mass or radius, we would retain the agreement in the mass-metallicity plane whilst getting a better match between model and data in the [Z/H]-$\Phi$ plane. We note, however, that this is a smaller increase in metallicity due to slow quenching than proposed by \citetalias{Peng:2015} and \citetalias{Trussler:2020}.

\subsection{Is the [Z/H]-$\Phi$ relation constant with redshift?}
\label{sec:validity_of_constant_Z_Phi_relation}

In our model, we made the simplifying assumption that the relationship between [Z/H] and $\Phi$ measured in Section \ref{sec:results_section} is the same at all redshifts. Since the relationship between [Z/H] and $\Phi$ has not been studied in a comparable manner at higher redshifts, assuming that it is constant over cosmic time is the simplest assumption to make without complicating our toy model by parameterising its evolution. As we have shown, making this assumption combined with a specific model of quenching, allows us to recover the [Z/H]-M$_*$ and [Z/H]-$\Phi$ relations of our sample.

However, the assumption of a constant relationship between [Z/H] and $\Phi$ is unlikely to be correct, as the following simple argument below shows. At a given mass, galaxies at high redshift are smaller than those at $z=0$ \citep[e.g.][]{VdW:2014}. This means that these high-redshift galaxies would have a larger value of $\Phi$ than their mass-matched counterparts at redshift zero. We also know that the mass-metallicity relationship evolves with redshift, for both star-forming \citep[e.g.][]{Cullen:2019, Kashino:2022} and quiescent galaxies \citep[e.g.][]{Leethochawalit:2018, Leethochawalit:2019}, with high redshift galaxies being more metal poor than today. Taken together, these facts imply that the relationship between [Z/H] and $\Phi$ must be shifted both downwards and to the right of the $z=0$ result from Section \ref{sec:results_section}. 

It is beyond the scope of this paper to further complicate our simple model by including evolution of the [Z/H]-$\Phi$ relation over cosmic time. We encourage simulators to study this relationship in their work (e.g. by studying [Z/H] in cosmological simulations; see e.g \citealt{Guo:2016}, \citealt{DeRossi:2017}, or use semi-analytic models which track stellar metallicity (e.g. SHARK;  \citealt{Lagos:2018}) to investigate whether they recover the results found in this work.

A key point to make here is that chemical evolution (i.e. an increase in metallicity) during quenching is intentionally ignored in the model, whilst on the other hand it is clear that that such  evolution will undoubtedly take place for the vast majority of galaxies (those which quench on timescales longer than a few Myrs). 

The larger the evolution in the [Z/H]-$\Phi$ relation (i.e. the greater departure from the assumption in this work), the greater the need for galaxies to increase their metallicity via chemical evolution during quenching, as discussed by \protect\citetalias{Peng:2015} and \protect\citetalias{Trussler:2020}. A key topic for future work will therefore be to measure the stellar metallicity, mass and size of galaxies at intermediate and high redshift in order to pin down this evolution (see e.g. \citealt{Barone:2021}).

\subsection{Comparison with previous work}

A number of previous studies have investigated quenching on populations of galaxies based on their evolution in M$_{*}$ and $r_e$. These studies tend to fall into two categories: those which have galaxies evolving in the mass size plane before quenching at fixed velocity dispersion, and those which find that galaxies quench at fixed surface density (M$_*/r_e^2$). \cite{vanderWel:2009}, \cite{Cappellari:2013}, \cite{vanDokkum:2015} and  \cite{Haines:2017} can be broadly grouped into the former category, whilst \cite{Tacchella:2015} and \cite{Barro:2017} and \cite{Barone:2021} belong to the latter. 

\cite{Chen:2020} take this area of modelling one step further and consider the evolution of galaxies in the four dimensional space of central surface density ($\Sigma_1$), M$_*$, $r_e$ and central black hole mass, M$_{\mathrm{\bullet}}$. Their work used straight lines in the $\Sigma_1$-M$_*$ plane which evolves with redshift as a quenching boundary, which they translated to the mass-size plane using the Se\'rsic indices of quiescent galaxies. Their key insight is showing that this quenching boundary in the $\Sigma_1$-M$_*$ plane can be related to the total energy deposited into the galaxy halo by the central black hole. Using such a model, \cite{Chen:2020} recover an impressive number of correlations and properties of the galaxy population in the local Universe.

Our study is the first to include stellar metallicity in a quenching model based on mass and size. We also use an "in-between" quenching boundary of M$_{*}/r_{e}^{3/2}$, instead of $M/r_{e}$ or $M/r_{e}^{2}$, since this gives the best match to our observational data.

We investigated a number of different boundaries for the regions where galaxies have a non-zero probability of quenching during their next timestep.  A quenching boundary at fixed mass leads to a vertical boundary between quenched and star-forming galaxies in the mass-metallicity plane, and cannot reproduce the fact that quiescent and star-forming galaxies of the same mass are observed. It would also not reproduce the mass-size plane we observe, which has a sloped line separating star-forming and passive galaxies. Similarly, a quenching boundary at fixed M/$r$ is unsatisfactory, since it tends to lead to a hard vertical line dividing quenched and star-forming galaxies in the [Z/H]-$\Phi$ plane. This is not observed in the data, where there is a small overlap between the two populations at $\Phi\approx10.2$. We also find that a fixed M/$r$ boundary leads to a steep slope in the [Z/H]-M$_*$ plane for the quiescent galaxies, as well as too few quenched galaxies at low mass. This last point means that the quenched fractions as a function of mass in Figure \ref{fig:quenching_f_mass_investigation} would not be a good match. On the other hand, quenching at a fixed surface density leads to too many low-mass quiescent centrals and a very flat slope in the [Z/H]-M$_*$ plane. We therefore concluded that a compromise worked best; a line of fixed \lmstar - $\frac{3}{2}\log_{10}(r_e/\mathrm{kpc})$, which has a slope somewhere between these two lines. 

We should stress that this exercise is designed to show that a sensible model using instantaneous quenching can be used to model galaxies in the [Z/H]-M$_*$ plane, and that modelling stellar metallicities in this way can be accomplished by using the relationship between [Z/H] and $\Phi$. We do not aim to conclude that the quenching boundaries in Figure \ref{fig:quenching_model_probability_mass_size_plane} supersede all others, or that this parameterisation of quenching should be taken as exactly what happens in the Universe. It is clear that the parameter space of possible such quenching models is very large, with the possibility of including time varying quenching strengths, different functional forms of quenching probability (e.g. see \citealt{Peng:2010} and \citealt{Lilly:2016}) and/or the inclusion of galaxy mergers and environmental-based quenching. Expanding our simple model in these ways is a topic for future work.

\begin{figure*}
    \centering
    \includegraphics[width=\textwidth]{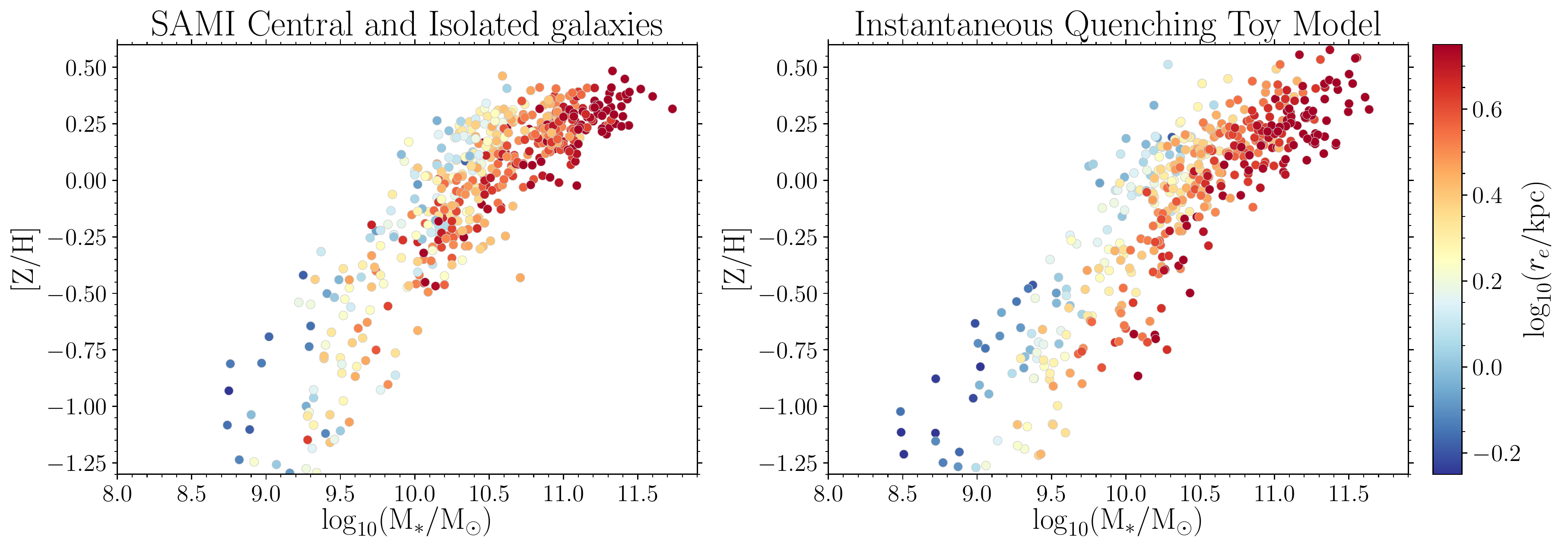}
    \caption{Plots of the mass-metallicity relationship for our SAMI galaxies (left) and simulated objects (right), with each point coloured by its effective radius. This figure emphasises that for two galaxies with the same mass, the one with the smaller radius will tend to have the higher metallicity, regardless of whether they are star-forming or quenched.}
    \label{fig:mass_zh_radius}
\end{figure*}

\subsection{Implications}

We find that our simple model of galaxy quenching which depends on both the mass and size of a galaxy can recover the difference in metallicity between star-forming and quiescent galaxies at fixed mass, implying that such a difference alone cannot be used as evidence of slow quenching processes. This is not to say that processes which remove a galaxy's halo of metal-poor gas are not important quenching pathways; evidence for disk strangulation (and its more extreme cousin ram-pressure stripping) has been found in the local Universe (e.g. \citealt{Poggianti:2017, Owers:2019} and see \citealt{Cortese:2021} for a review), at higher redshifts \citep[e.g.][]{Maier:2016, Maier:2019, Vaughan:2020} and in simulations \cite[e.g.][]{DeRossi:2015}. 

Our model also implies that merging processes are not necessarily important in setting the slopes of the mass-size, [Z/H]-$\mathrm{M}_{*}$ and [Z/H]-$\Phi$ relations, since we recover these in a model that does not contain any galaxy interactions explicitly\footnote{It should be noted that we have sampled our simulated galaxies to follow the same mass-size relation as the observed SAMI galaxies, however. If mergers play a role in shaping the observed mass-size relation, they will be implicitly included in our model.}. The fact that our toy model of galaxy evolution recovers a number of important results without explicit major merging is perhaps not too surprising, since simulations have suggested that the rate of growth due to major mergers (mass ratio greater than 1:10) contribute only 20\% of a galaxy's overall mass growth \citep{Wang:2011, LHuillier:2012}. Whilst major mergers are more important in high-density environments, it is clear that gas accretion from the cosmic web and in-situ star formation is the predominant mass-growth channel for most galaxies in the Universe.

\subsubsection{What leads to the difference in [Z/H] between passive and star-forming galaxies?}
\label{sec:offset_in_MZ_relation}

By tracing the metallicity, mass and size history of the galaxies in our toy model, we can investigate the cause of metallicity difference between quenched and star-forming galaxies at fixed mass.

\begin{figure}
    \centering
    \includegraphics[width=\columnwidth]{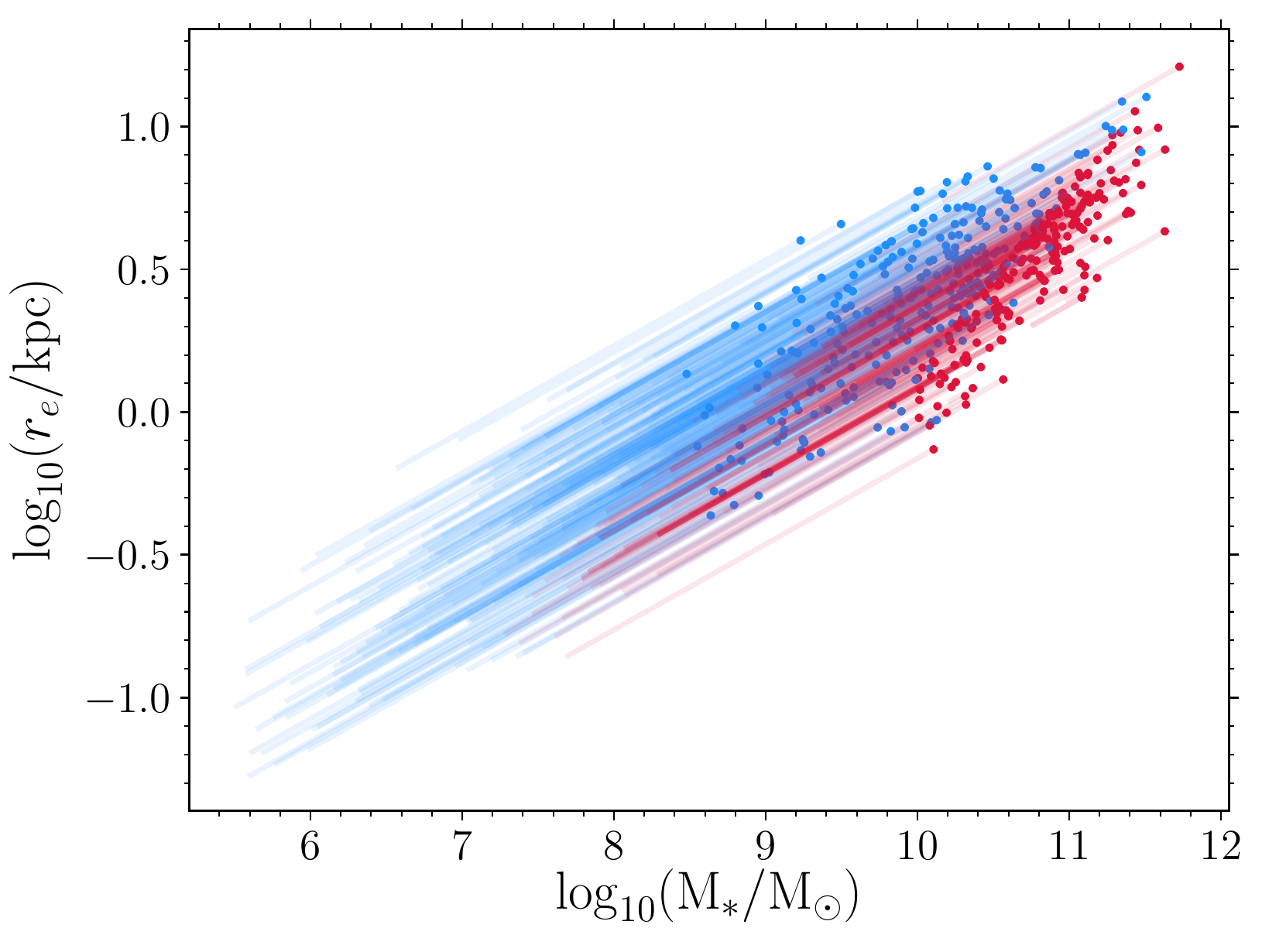}
    \caption{Tracks in the mass-size plane for our simulated galaxies. This plot emphasises that our model galaxies move in parallel in the mass-size plane, and that galaxies which are quiescent at redshift zero have always been amongst the smallest objects for their mass in a population.}
    \label{fig:mass_size_tracks}
\end{figure}

Figure \ref{fig:mass_size_tracks} shows the mass-size plane of simulated galaxies which have been chosen to match our observational sample\footnote{We note that here we are plotting the mass and radius values from the simulation without any "observational" uncertainties applied, unlike in the bottom panel of Figure \ref{fig:quenching_toy_model}}. For each model galaxy, we plot its trajectory in the mass-size plane from the time it enters our model (i.e. from the starting mass, size and formation redshift it is assigned). Galaxies which are passive at redshift zero are coloured in red, whilst galaxies which are still star-forming are blue. 

At any given mass, quenched galaxies tend to have smaller radii than galaxies which are still star-forming \citep[e.g.][]{VdW:2014}. Figure \ref{fig:mass_size_tracks} reinforces that in our model this is (by construction) true at all times in the past: galaxies which end up passive at redshift zero are (nearly) always smaller than those which end up to be star-forming. This is a simple consequence of the modelling assumptions we have made. Equation \ref{eqtn:dm_dr} implies that all galaxies follow parallel tracks in the mass-size plane whilst they are star-forming, and so the smallest galaxies in a population remain the smallest.

This fact impacts the final metallicities of the galaxy population through our assumption that [Z/H] traces $\Phi = \log_{10}\left(\frac{M_*}{M_{\odot}} \right) - \log_{10}\left(\frac{r_e}{\mathrm{kpc}} \right)$ at all times in a galaxy's history (and see also \cite{Barone:2021} who show that [Z/H] traces $\Phi$ in their sample of galaxies at $z=0.76$). At any given point, galaxies with small radii for their mass will have larger values of $\Phi$, and so will tend to have a larger value of [Z/H]. This is not only true when comparing passive and star-forming objects; as shown in Figure \ref{fig:mass_zh_radius}, smaller galaxies tend to have higher stellar metallicities even in locations of the diagram where everything is star-forming. Our model implies that this residual size dependence is baked in from a galaxy's formation: objects with higher metallicities at fixed mass have always been the smallest objects for their mass in a population, and it is this fact which leads to their location in the mass-metallicity plane today (e.g. as seen in Figure \ref{fig:mass_zh_radius}). 

Related to this idea is our quenching prescription: galaxies quench once they cross a diagonal boundary in the mass-size plane. By design, this boundary will tend to quench galaxies which are small for their mass and "freeze" their values of mass, size and metallicity in place. Since this prescription does not allow for galaxies to increase their metallicity after quenching, however, it does not contribute to the offset in metallicity at fixed mass between star-forming and passive galaxies (unlike the slow-quenching described in \citetalias{Peng:2015} and \citetalias{Trussler:2020}). 

\section{Conclusions}
\label{sec:conclusions}

This work builds on the studies of \citet{Barone:2018}, \citet{D'Eugenio:2018} and \citet{Barone:2020} by investigating the correlation between stellar metallicity ([Z/H]) and $\Phi = \log_{10}\left(\frac{M_*}{M_{\odot}} \right) - \log_{10}\left(\frac{r_e}{\mathrm{kpc}} \right)$ for a homogeneously-observed sample of galaxies. We use the SAMI galaxy survey to measure central metallicity values for 1363 galaxies which also have robust stellar mass, half-light radius and star-formation rate measurements, and build a simple toy model of galaxy evolution which has galaxies evolving in mass, size and [Z/H] to explain our findings. Galaxies in this model have a probability $p_{\mathrm{quench}}$ to undergo nearly instantaneous quenching (from main-sequence to fully passive in under 100 Myrs). 

Our findings are as follows:

\begin{enumerate}
    \item We recover the well-known offset between the metallicity of star-forming and quiescent galaxies at fixed mass, with our results in very good agreement with the previous work of \citet{Peng:2015} and \citet{Trussler:2020}. 
    \item We show for the first time that star-forming, passive and intermediate ("green valley") galaxies form a single sequence in the  [Z/H]-$\Phi$ relation, with passive galaxies preferentially being found to the right of the plot (at larger values of $\Phi$).
    \item The difference in metallicity between passive and star-forming galaxies is smaller at fixed $\Phi$ than at fixed \lmstar. At fixed $\Phi$, $\Delta$[Z/H] is $\approx$ 0.1 dex for all values of $\Phi$. This is smaller than or equal to the difference between star-forming and passive galaxies at fixed mass, which ranges from $\approx$0.1 dex at high masses and up to $\approx$0.35 dex below \lmstar=10.
        \item In our toy model of galaxy evolution with instantaneous quenching, choosing a quenching probability which depends on a galaxy's mass \textit{and size} allows us to quantitatively recover the slope of the [Z/H]-$\mathrm{M}_{*}$ relation for our sample, as well as account for the offset in metallicity between star-forming and passive galaxies at fixed mass. We therefore conclude that this offset is not necessarily the result of slow quenching, in contrast to the studies of \citet{Peng:2015} and \citet{Trussler:2020}.
    \item We do find that our toy model slightly underpredicts the difference in metallicity between star-forming and passive galaxies at fixed $\Phi$ by $\approx$ 0.05 dex. If galaxies increase in metallicity by $\approx$ 0.03--0.05 dex whilst quenching (e.g. during the processes discussed in \citetalias{Peng:2015} and \citetalias{Trussler:2020}), our model would retain the good agreement in the mass-metallicity plane whilst giving a closer match in the [Z/H]-$\Phi$ plane. We note, however, that this amount of slow quenching is much less than \citetalias{Peng:2015} and \citetalias{Trussler:2020} propose.
\end{enumerate}

This study adds to the collection of work which shows the importance of accounting for galaxy size when investigating stellar populations, and not only considering stellar mass. Future work to investigate the [Z/H] properties of galaxies at higher redshifts (e.g in the forthcoming MAGPI survey; \citealt{Foster:2020}), as well as in nearby low-mass galaxies and comparing to cosmological simulations, will be invaluable to understand the relationship between galaxy mass, size and stellar metallicity.

\section*{Acknowledgements}

We would like to thank the anonymous referee for their detailed and thorough response which greatly improved this work.
This research was conducted using the freely available \textsc{Python} programming language \citep{vanRossum:1995} and the \textsc{IPython} extension \citep{IPython}. Our analysis made use of the \textsc{Numpy} \citep{Numpy}, \textsc{Scipy} \citep{Scipy}, \textsc{Astropy} \citep{Astropy}, \textsc{matplotlib} \citep{Hunter:2007}, and \textsc{Pandas} \citep{pandas} packages. We also use the \textsc{Stan} probabilistic programming language \citep{stan} via its python extension, \textsc{Pystan}. We also use the \textsc{snakemake} workflow management system \citep{snakemake} to ensure our analysis is reproducible and transparent. We make extensive use of NASA’s Astrophysics Data System.

The SAMI Galaxy Survey is based on observations made at the Anglo-Australian Telescope. The Sydney-AAO Multi-object Integral field spectrograph (SAMI) was developed jointly by the University of Sydney and the Australian Astronomical Observatory. The SAMI input catalogue is based on data taken from the Sloan Digital Sky Survey, the GAMA Survey and the VST ATLAS Survey. The SAMI Galaxy Survey is supported by the Australian Research Council Centre of Excellence for All Sky Astrophysics in 3 Dimensions (ASTRO 3D), through project number CE170100013, the Australian Research Council Centre of Excellence for All-sky Astrophysics (CAASTRO), through project number CE110001020, and other participating institutions. The SAMI Galaxy Survey website is http://sami-survey.org/. LC is the recipient of an Australian Research Council Future Fellowship (FT180100066) funded by the Australian Government. FDE acknowledges funding through the H2020 ERC Consolidator Grant 683184, the ERC Advanced grant 695671 "QUENCH'' and support by the Science and Technology Facilities Council (STFC). TMB is supported by an Australian Government Research Training Program Scholarship. JJB acknowledges support of an Australian Research Council Future Fellowship (FT180100231). JvdS acknowledges support of an Australian Research Council Discovery Early Career Research Award (project number DE200100461) funded by the Australian Government. NS acknowledges support of an Australian Research Council Discovery Early Career Research Award (project number DE190100375) funded by the Australian Government and a University of Sydney Postdoctoral Research Fellowship. SB acknowledges funding support from the Australian Research Council through a Future Fellowship (FT140101166). MSO acknowledges the funding support from the Australian Research Council through a Future Fellowship (FT140100255). JBH is supported by an ARC Laureate Fellowship FL140100278. The SAMI instrument was funded by Bland-Hawthorn's former Federation Fellowship FF0776384, an ARC LIEF grant LE130100198 (PI Bland-Hawthorn) and funding from the Anglo-Australian Observatory.

We acknowledge the traditional custodians of the land on which the AAT stands, the Gamilaraay people, and pay our respects to their elders past and present.

\section*{Data Availability}

The observations used in this work have been released as part of the SAMI Data Release 3 \citep{Croom:2021}. The metallicity measurements used in this article will be shared on reasonable request to the corresponding author.

\bibliographystyle{mnras}
\bibliography{bibliography}

\appendix

\section{Uncertainties in [Z/H]}
\label{sec:uncertainties}

We require an estimate of the uncertainty in our [Z/H] measurements from each individual Voronoi-binned spectrum. However, running a bootstrap analysis using 50 bootstraps on all 78,531 spectra would take approximately 2.5 years of CPU time and be an unnecessary waste of supercomputer resources and energy. Instead, we bootstrap a subsample of $\approx$ $10^3$ spectra and use their measured uncertainties to build a model to predict the uncertainty of the remaining spectra.

We carefully select this subsample to span the entire range of S/N and [Z/H] values in our data. We note that selecting a random sample of spectra to bootstrap would \textit{not} be appropriate. Since the majority of our spectra are around solar metallicity, bootstapping a random sample would necessarily lead to a poor prediction of the uncertainties for low-metallicity spectra. 

\begin{figure}
    \centering
    \includegraphics[width=\columnwidth]{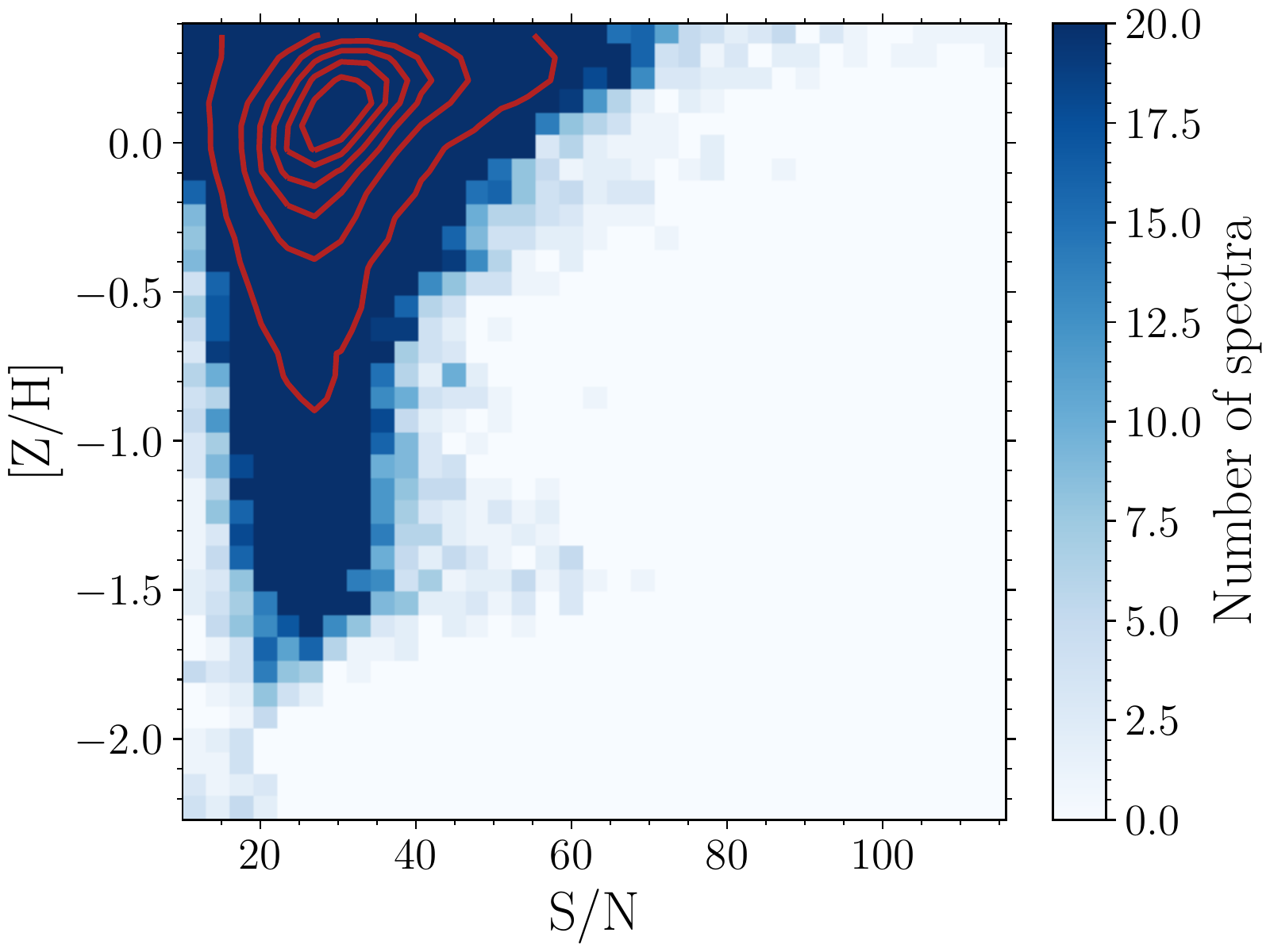}
    \caption{The distribution of [Z/H] and S/N values of the spectra in our sample. The colormap shows the number of spectra from each grid cell chosen for bootstrapping. We select all spectra in each cell unless there are more than 20, in which case we choose a random subsample of 20 spectra. The red contours show the distribution of spectra in the full sample.}
    \label{fig:bootstrapping_2d_hist}
\end{figure}

We divide the S/N and [Z/H] plane of our spectra into a 30x30 grid. We select 20 spectra at random from each grid cell, and all the spectra in a grid cell if there are fewer than 20. The two-dimensional histogram of the S/N and [Z/H] properties of the spectra selected for bootstrapping are shown in Figure \ref{fig:bootstrapping_2d_hist}, as well as the distribution of the entire sample. Our final bootstrapping sample contains 10,814 spectra. 

For each spectrum in the bootstrapping sample, we take the best fitting template from \textsc{pPXF} and add Gaussian random noise at each pixel. This random noise is scaled appropriately by the variance spectrum of the observed data, multiplied by the reduced $\chi^2$ value from the \textsc{pPXF} fit. We then re-fit this spectrum with \text{pPXF} and measure the stellar population parameters. This process is repeated 50 times per spectrum. 

We aim to predict the bootstrapped uncertainty, $\sigma_{[\mathrm{Z/H}]}$, of a spectrum from its S/N and [Z/H] values. To assess the success of our model, we hold back a "test set" of 20\% of the data (2163 spectra) and use the remaining 80\% to fit to. 

\begin{figure}
    \centering
    \includegraphics[width=\columnwidth]{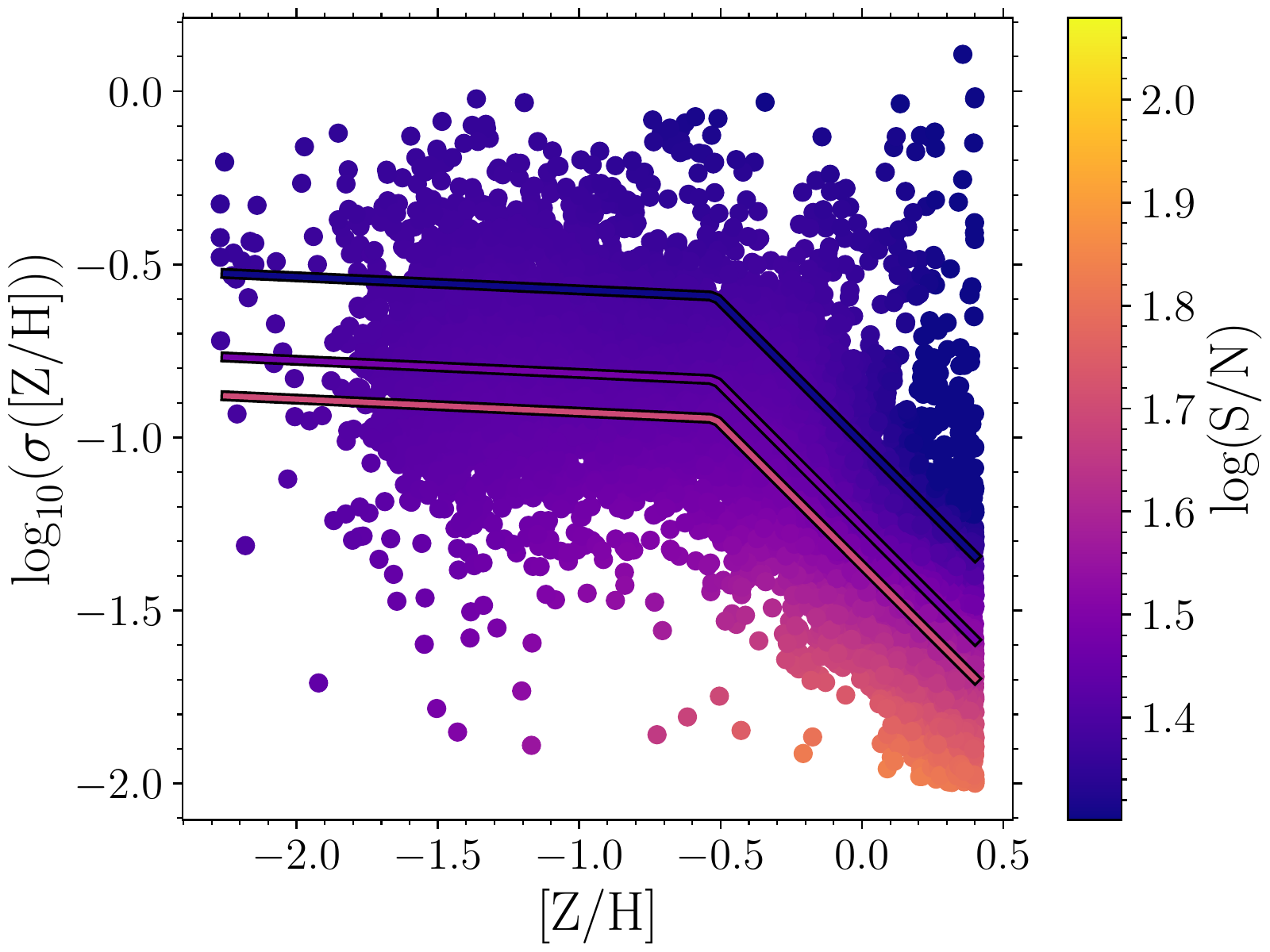}
    \caption{A plot of the bootstrapped uncertainty ($\log_{10}(\sigma_{[\mathrm{Z/H}]})$) against the metallicity of each spectrum. The colourbar corresponds to $\log_{10}(\mathrm{S/N})$ of the spectrum, and has been smoothed using the \texttt{loess2d} locally-weighted regression algorithm described in \protect\cite{Cappellari:2013b}. }
    \label{fig:sigma_against_ZH}
\end{figure}

As shown in Figure \ref{fig:sigma_against_ZH}, the dependence of $\log_{10}(\sigma_{[\mathrm{Z/H}]})$ with [Z/H] is roughly linear for spectra with [Z/H] above $-0.5$, but becomes approximately flat with [Z/H] below this value. As expected, higher S/N spectra have smaller uncertainties. We note that the colormap has been smoothed using the \texttt{loess2d} algorithm described in \citet{Cappellari:2013b}

We therefore capture this behaviour with a simple broken line model. Defining $y=\log_{10}(\sigma_{[\mathrm{Z/H}]})$ and $s=\log_{10}({\mathrm{S/N}})$, the model has 6 free parameters:

\begin{enumerate}
    \item $m_1$, the dependence of $y$ on $s$
    \item $z_b$, the breakpoint in the  $y$-[Z/H] relationship. 
    \item $m_2$, the dependence of $y$ on [Z/H] for [Z/H] $\geq z_b$
    \item $m_3$, the dependence of $y$ on [Z/H] for [Z/H] <$z_b$
    \item $a$, the intercept
    \item $\tau$, the intrinsic scatter around the relation
\end{enumerate} 

and is described by the following equation:

\begin{equation}
\label{eqtn:broken_line}
    y {} = 
    \begin{cases}
      a + m_{1}s + m_{2}\mathrm{[Z/H]} &\text{if           [Z/H] $\geq z_b$}\\
      a + m_{1}s + m_{3}\mathrm{[Z/H]}  + ( m_{3} - m_2)z_b & \text{if [Z/H] < $z_b$}
    \end{cases}
\end{equation}

Our results are summarised in Table \ref{tab:uncertainty_model_bestfit_params}. The standard deviation of the residuals between the metallicity uncertainty of the test set and the metallicity uncertainty predicted by the model (where both values are on the log scale) is 0.24 dex, corresponding to us being able to estimate the true uncertainty of a spectrum to better than a factor of 2. The standard deviation on the linear scale  (i.e. in terms of $\sigma_{[\mathrm{Z/H}]}$) is 0.09 dex. 

\begin{table*}
    \centering
    \begin{tabular}{cccccc}
         \toprule
         $a$ & $m_1$ & $m_2$ & $m_3$ & $z_b$ & $\tau$ \\
         \midrule
         $-0.09\pm0.03$& $-0.50\pm0.02$& $-0.03\pm0.01$ &  $-0.82\pm0.01$ & $-0.53\pm0.02$ & $0.24\pm0.01$\\
         \bottomrule
    \end{tabular}
    \caption{Best fitting parameters of the broken line model used to characterise the uncertainties described in Equation \ref{eqtn:broken_line}.}
    \label{tab:uncertainty_model_bestfit_params}
\end{table*}

\section{Using a different set of SSP models}
\label{sec:BPASS}

In this section, we investigate the affect of using a different set of SSP models to measure the stellar metallicity of our sample. We randomly select 130 galaxies from the full sample (approximately 10\%, corresponding to 5939 Voronoi bins) and use \ppxf{} to fit version 2.2.1 of the Binary Population and Spectral Synthesis (BPASS) models \citep{Eldridge:2017,Stanway:2018,Byrne:2022} to their spectra. These BPASS templates have an [$\alpha/\mathrm{Fe}$] ratio of 0.0 and use a Chabrier initial mass function \citep{Chabrier:2003}. We also only use the "single star" templates, which do not include the effect of binary stars on the final SSP spectra, to more closely match the MILES models. We perform the fit using 228 templates which form a grid in age and metallicity, spanning from 6 Myrs to 15.8 Gyrs in age and from -4.3 to +0.3 in [Z/H]. \ppxf{} is used to perform the fit, with the same settings as in Section \ref{sec:stellar_pop_measurements}.

\begin{figure*}
    \centering
    \includegraphics[width=\textwidth]{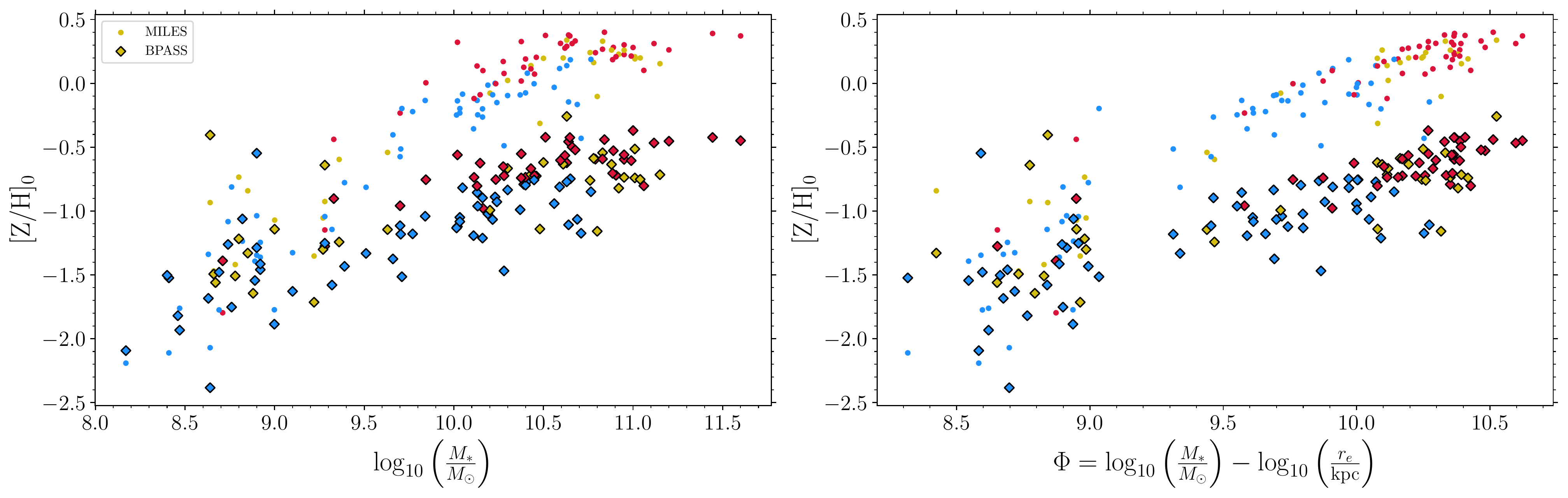}
    \caption{A comparison of the mass-metallicity and $\Phi$-metallicity relations for 130 galaxies in our sample using the MILES models (circles) and BPASS models (diamonds). The colour of each point denotes whether the galaxy is star-forming (blue), passive (red) or "green-valley"/intermediate (yellow), as defined in Section \ref{sec:stellar_pop_measurements}. We see that although the BPASS measurements are offset to lower metallicities, especially for galaxies with high values of mass or $\Phi$, the same overall trends remain in both cases.}
    \label{fig:bpass_vs_MILES}
\end{figure*}

Figure \ref{fig:bpass_vs_MILES} shows the derived mass-metallicity and $\Phi$-metallicity relations. We see that measurements with the BPASS models are offset to lower values of metallicity, with the difference being as large as 1 dex in the most discrepant cases at high values of mass or $\Phi$.

The fact that two different SSP models do not precisely agree on the absolute values of metallicity in a spectrum is not particularly surprising, since finding differences in the recovery of stellar population parameters when using different SSP models is well known \citep[e.g.][]{GonzlezDelgado:2010, Fan:2016, Baldwin:2018, Ge:2019}. Whilst a full investigation of the reasons behind the different derived metallicities from the MILES and BPASS models is beyond the scope of this paper, we note the large number of different ingredients between the two; the models use different isochrones (the BaSTI isochrones for MILES, a custom stellar evolutionary prescription described in \cite{Eldridge:2017} for BPASS), different stellar libraries (empirical stellar spectra for MILES, theoretical model atmospheres for BPASS) and have a different high-mass cutoff for the assumed IMF (100\Msun{} for MILES compared to 300\Msun{} for BPASS). We also note the discussion in \cite{Eldridge:2017} regarding the difficulty in reproducing some aspects of old stellar populations using the BPASS models, since they were primarily designed to study populations at less than 1Gyr. We therefore point the interested reader to the papers describing each SSP library for further details; \cite{Vazdekis:2015} and \cite{Eldridge:2017}. 

Having said that, the overall trends which we are interested in remain the same for the two sets of measurements. In each case, we see that there is an offset in metallicity between star-forming and passive galaxies at fixed mass, and in both cases we see a tighter correlation between [Z/H] and $\Phi$ than between [Z/H] and mass. We therefore find that our conclusions regarding [Z/H], \lmstar and $\Phi$ are unchanged regardless whether we use the MILES or BPASS stellar population models.

\section{A comparison between [Z/H]$_{0}$ and [Z/H]$_{r_e}$}
\label{sec:central_vs_1re_comparison}

\begin{figure*}
    \centering
    \includegraphics[width=\textwidth]{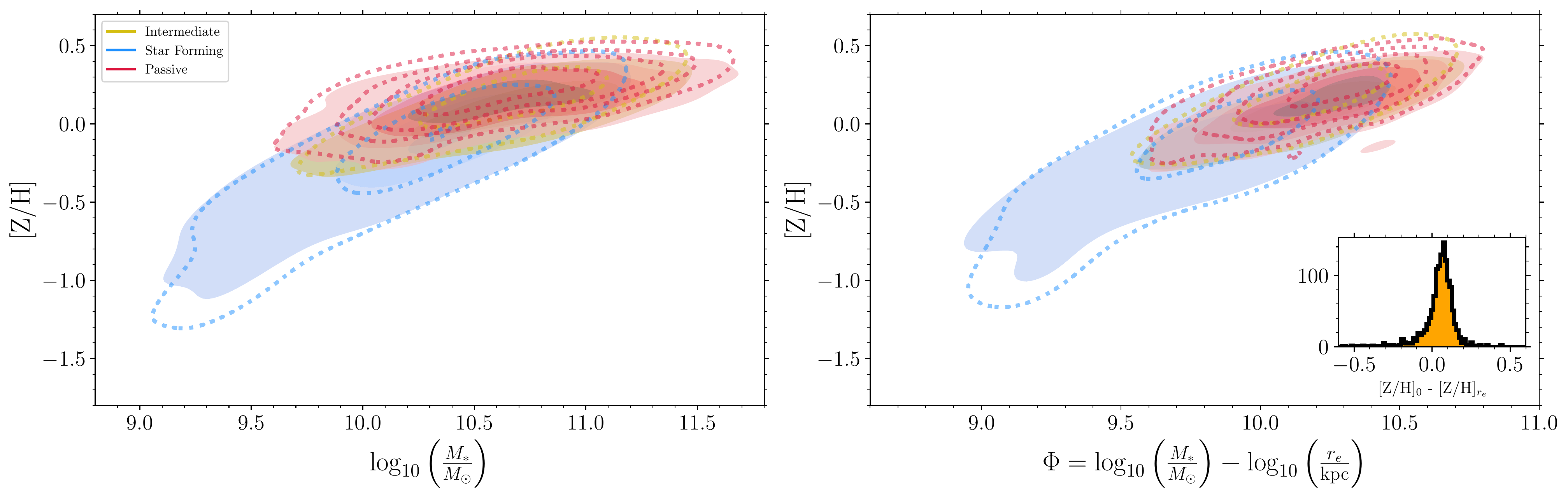}
    \caption{A comparison of the mass-metallicity and $\Phi$-metallicity relations using the central metallicity of each galaxy from the model described in Section \ref{sec:more_complicated} ([Z/H]$_{0}$); and the metallicity derived from the 1 $r_e$ spectrum of each galaxy ([Z/H]$_{r_{e}}$). Shaded density plots correspond to the metallicities from 1 $r_e$ and dotted contour lines show central metallicities. The colours denote the locations of star-forming (blue); passive (red); or intermediate (yellow) galaxies. The inset panel to the right shows the distribution of [Z/H]$_{0}$ - [Z/H]$_{r_{e}}$, which has a mean of 0.04 and a standard deviation of 0.15.}
    \label{fig:1_re_central_comparison}
\end{figure*}

A comparison between our results using the central metallicity and the metallicity measured from the 1 $r_e$ spectrum of each galaxy is shown in Figure \ref{fig:1_re_central_comparison}. This choice makes only a marginal difference to the [Z/H] values for each galaxy, leaving our conclusions unchanged. On average, the central metallicity is 0.04 dex larger than the metallicity from within 1 $r_e$. The standard deviation of the difference is 0.15 dex, and a histogram is shown in the inset panel of Figure \ref{fig:1_re_central_comparison}.

\section{Velocity dispersion or $\Phi$?}
\label{sec:sigma_vs_phi}

\begin{figure}
    \centering
    \includegraphics[width=\columnwidth]{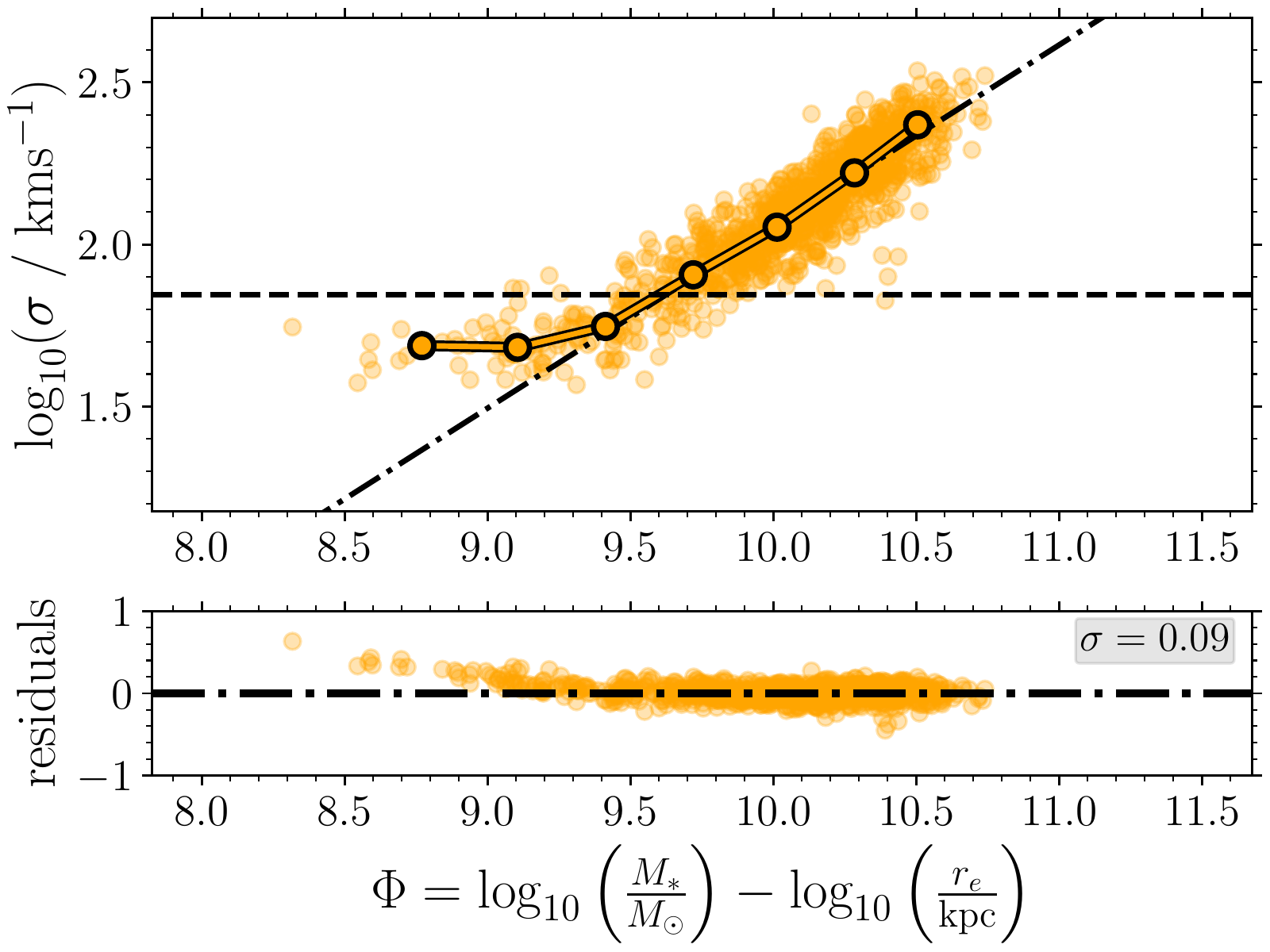}
    \caption{The logarithm of the velocity dispersion of the central Voronoi bin ($\sigma_0$) for each galaxy against the galaxy's value of $\Phi = \log_{10}\left(\frac{M_*}{M_{\odot}} \right) - \log_{10}\left(\frac{r_e}{\mathrm{kpc}} \right)$. The expected linear relationship is clear for galaxies with $\sigma_0$ above the SAMI blue-arm instrumental resolution of $\approx70$ kms$^{-1}$ (shown by the dashed line).}
    \label{fig:sigma_vs_potential}
\end{figure}

Figure \ref{fig:sigma_vs_potential} shows the relationship between the logarithm of the central velocity dispersion in each galaxy and its value of  $\Phi = \log_{10}\left(\frac{M_*}{M_{\odot}} \right) - \log_{10}\left(\frac{r_e}{\mathrm{kpc}} \right)$. The linear relationship is evident above the SAMI instrumental resolution of $\approx70$ kms$^{-1}$. As discussed in Section \ref{sec:discussion}, we use $\Phi$ in this work to allow us to include galaxies in this work with $\Phi < 9.5$, without the issue of measuring $\sigma$ close to the instrumental resolution.

\section{A comparison between our straight line models}
\label{sec:hierarchical_normal_comparison}
We now present a comparison between the results of the simple straight line fit described in Section \ref{sec:simple_model} and the more complicated hierarchical model account for censored data described in Section \ref{sec:more_complicated}.

\begin{figure}
    \centering
    \includegraphics[width=\columnwidth]{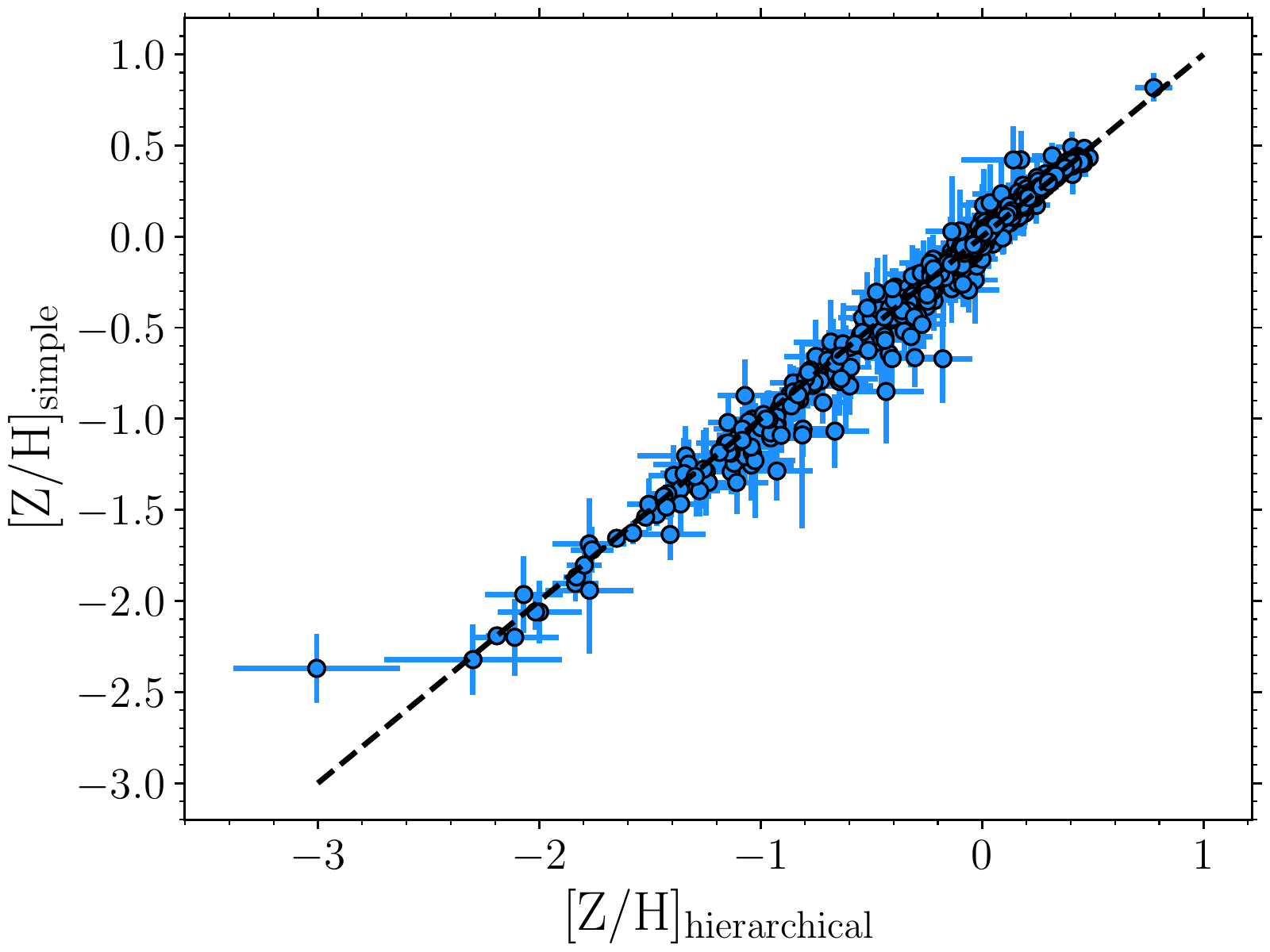}
    \caption{A comparison of the central metallicity value (the straight line intercept) for each galaxy derived from the hierarchical model accounting for censored data (on the $x$ axis) and the simple straight line fit ($y$ axis). We see a very good agreement between the two. The difference between the two has mean 0.006 dex and standard deviation 0.05 dex.}
    \label{fig:ZH_comparison_normal_hierarchical}
\end{figure}

Figure \ref{fig:ZH_comparison_normal_hierarchical} shows the central metallicity value for each galaxy derived from the hierarchical model accounting for censored data plotted against the value derived from the simple straight line fit. We find a very good agreement between the two, with their difference having mean 0.006 dex and standard deviation 0.05 dex.

\bsp	\label{lastpage}
\end{document}